\documentclass[twocolumn,twocolappendix]{aastex63}

\usepackage{xspace}
\usepackage{chngcntr}
\usepackage[multiple]{footmisc}
\usepackage{threeparttable}

\newcommand\chandra{\textit{Chandra}\xspace}
\newcommand{\ciao}{\textit{CIAO}\xspace}
\newcommand{\marx}{\textit{MARX}\xspace}
\newcommand{\saotrace}{\textit{SAOTrace}\xspace}

\newcommand{\msol}{\ensuremath{\mbox{$\mathrm{M}_{\sun}$}}\xspace}

\newcommand{\xspec}{\texttt{Xspec}\xspace}

\graphicspath{{./}{figures/}}

\received{2020.07.14}
\accepted{2020.10.10}

\shorttitle{Putative CCO in 1E 0102.2-7219}
\shortauthors{Xi et al.}

\begin{document}

\title{On the X-ray Properties of the Putative Central Compact Object in 1E 0102.2-7219}

\correspondingauthor{Xi Long}
\email{xi.long@cfa.harvard.edu}

\author{Xi Long \href{https://orcid.org/0000-0003-3350-1832}{\includegraphics[scale=0.01]{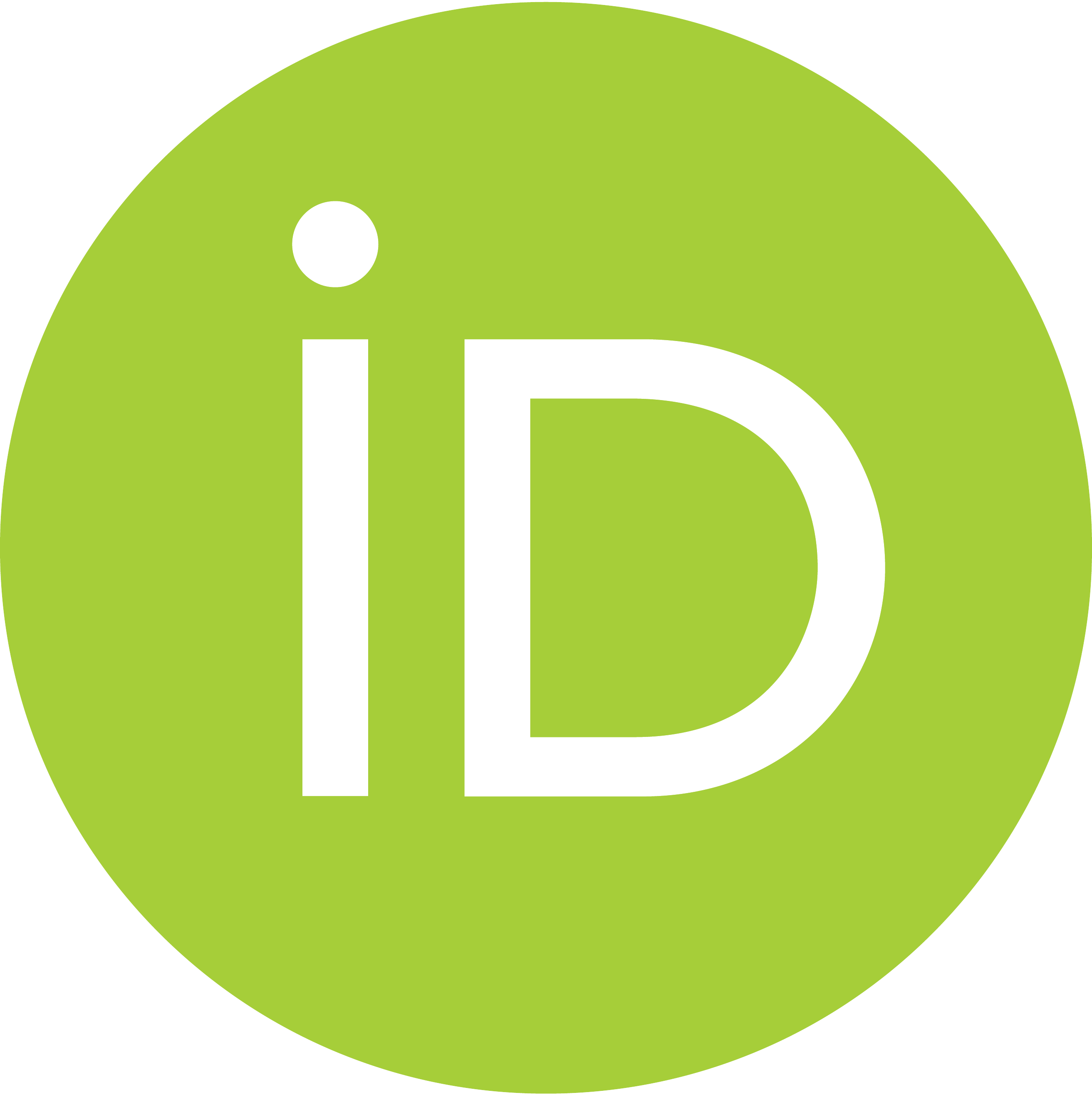}}}
\affiliation{Center for Astrophysics --- Harvard \& Smithsonian, 60 Garden St., Cambridge, MA 02138, USA; xi.long@cfa.harvard.edu,tgaetz@cfa.harvard.edu, pplucinsky@cfa.harvard.edu}
\affiliation{Purple Mountain Observatory, CAS, Nanjing 210023, P.R.China}
\affiliation{Key Laboratory of Dark Matter and Space Astronomy, CAS, Nanjing 210023, People’s Republic of China}
\author{Terrance J. Gaetz \href{https://orcid.org/0000-0002-5115-1533}{\includegraphics[scale=0.01]{orcid.pdf}}}
\affiliation{Center for Astrophysics --- Harvard \& Smithsonian, 60 Garden St., Cambridge, MA 02138, USA; xi.long@cfa.harvard.edu,tgaetz@cfa.harvard.edu, pplucinsky@cfa.harvard.edu}
\author{Paul P. Plucinsky \href{https://orcid.org/0000-0003-1415-5823}{\includegraphics[scale=0.01]{orcid.pdf}}}
\affiliation{Center for Astrophysics --- Harvard \& Smithsonian, 60 Garden St., Cambridge, MA 02138, USA; xi.long@cfa.harvard.edu,tgaetz@cfa.harvard.edu, pplucinsky@cfa.harvard.edu}

\begin{abstract}
We have analyzed the archival Chandra X-ray Observatory observations of the compact feature in the Small Magellanic Cloud supernova remnant (SNR) 1E 0102.2-7219 which has recently been suggested to be the Central Compact Object remaining after the supernova explosion. In our analysis, we have used appropriate, time-dependent responses for each of the archival observations, modeled the background instead of subtracting it, and have fit unbinned spectra to preserve the maximal spectral information.  The spectrum of this feature is similar to the spectrum of the surrounding regions which have significantly enhanced abundances of O, Ne, \& Mg. 
We find that the previously suggested blackbody model is inconsistent with the data as Monte Carlo simulations indicate that more than 99\% of the simulated data sets have a test statistic value lower than that of the data. The spectrum is described adequately by a non-equilibrium ionization thermal model with two classes of models that fit the data equally well. One class of models has a temperature of $kT\sim0.79$~keV, an ionization timescale of $\sim3\times10^{11}\,\mathrm{cm}^{-3}\mathrm{s}$, and marginal evidence for enhanced abundances of O and Ne and the other has a temperature of $kT\sim0.91$~keV, an ionization timescale of  $\sim7\times10^{10}\,\mathrm{cm}^{-3}\mathrm{s}$, and abundances consistent with local interstellar medium values. We also performed an image analysis and find that the spatial distribution of the counts is not consistent with that of a point source. The hypothesis of a point source distribution can be rejected at the 99.9\% confidence level. Therefore this compact feature is most likely a knot of O and Ne rich ejecta associated with the reverse shock.
\end{abstract}


\section{Introduction}\label{sec:intro}

The collapse of a massive star during a supernova explosion may result in the formation of a compact object, either a neutron star (NS) or a black hole (BH). There are numerous examples of NSs that are associated with supernova remnants (SNRs) (see \citealt{kaspi2002}) supporting the idea that NSs are one of the possible byproducts of supernovae (SNe).  NSs make up a diverse population with various classes of objects, see \citet{harding2013} for a review. The classes include the rotation-powered pulsars, the isolated NSs,  accreting NSs in binary systems (which are further sub-divided into low-mass X-ray binaries and high-mass X-ray binaries), and the so-called magnetars, NSs with magnetic fields of $10^{14} - 10^{15}$~G. A new class of NSs was identified soon after the launch of the {\em Chandra X-ray Observatory} (\chandra). An X-ray point source was detected near the center of the Cassiopeia A (Cas~A) SNR \citep{tananbaum1999} and was suggested to be the compact object remaining after the SN explosion. This source was characterized by soft and steady thermal emission (equivalent to a blackbody temperature of $\mathrm{kT}\sim 0.4$~keV), lack of any extended emission that might be interpreted as a nebula, and no counterpart at other wavelengths \citep{pavlov2000}.  Since the discovery of this object, similar objects have been detected in other SNRs (all located near the center of the SNR) and this class has been given the name of Central Compact Objects (CCOs). There are now approximately eleven CCOs and CCO candidates as summarized by \cite{gotthelf2013}. See \cite{deluca2017} for a recent review article.
Pulsations have been detected for three of the CCOs strengthening the connection to the other classes of NSs, see \cite{zavlin2000} and \cite{gotthelf2013}.  The first timing glitch in a CCO pulsar was reported in \cite{gotthelf2018}. The magnetic field strength has been estimated to be $\sim 0.3-1.0\times10^{11}$~G for the three CCO pulsars based on the spin-down properties \citep{gotthelf2013}, which is two orders of magnitude lower than typical young pulsars and hence CCOs have been suggested to be ``anti-magnetars''.  It is not known if these objects were born with such a low value of the magnetic field or if the magnetic field has been buried by fallback accretion \citep{ho2011}.

The CCO in Cas A is perhaps the best studied as numerous observations have been made with \chandra\/ over the last 20 years. \cite{pavlov2009} found no evidence of pulsations or an extended nebula around the source and determined that the spectrum could not be fit well with a simple continuum model but required a NS atmosphere model. They concluded that the surface magnetic field is low ($\sim10^{11}$~G) and the temperature distribution is nonuniform on the surface of the NS. \cite{ho09} suggest the NS has a carbon atmosphere. A carbon atmosphere was also suggested for other CCOs by \cite{klochkov13}, \cite{klochkov16}, and \cite{Doroshenko18}.
\cite{heinke2010} reported an apparent cooling of the NS as the temperature declined by 4\% and the flux declined by 21\% over a nine year time interval.  \cite{posselt2018} claim that this apparent temperature decrease is not significant at the $3\sigma$ level and may have been affected by instrumental issues.
If rapid cooling were to be observed in the Cas A CCO it would have profound implications for the interior structure of NSs.

Given the many puzzles about the origin and evolution of CCOs, it is imperative to discover and study as many objects of this class as possible to constrain their properties.  All of the CCOs and CCO candidates are within the Galaxy given their relatively low luminosity ($\sim10^{33}~\mathrm{erg~s^{-1}}$ in the X-ray band). Recently, the first detection of an extragalactic CCO was claimed by \cite{vogt2018} in the SNR 1E\,0102.2-7219 (hereafter E0102) in the Small Magellanic Cloud (SMC).
An earlier effort by \cite{rutkowski2010} did not find conclusive evidence for a point source using the \chandra\/ data available through 2009.
It is expected that a compact object, either a NS or BH, formed in the explosion that produced E0102 given that the estimates of the progenitor mass range from 25\msol \citep{blair2000} to 40\msol \citep{alan2019}.
\cite{seitenzahl2018} argue that the progenitor was a Type~IIb SNe based on the detection of blueshifted and redshifted hydrogen with velocities similar to that measured for the other optical ejecta.
\cite{vogt2018} describe an enhancement in the X-ray emission surrounded by a ring-like structure in the optical recombination lines of \ion{Ne}{1} and \ion{O}{1}.  A crucial point in their argument that this object is indeed a NS is that the spectral distribution of the X-ray emission was claimed to be consistent with a blackbody spectrum.
Their analysis binned the spectral data into four broad spectral bands thereby reducing the sensitivity to line-like features, adopted instrument response files appropriate for 2017 to analyze data spanning the range from 2003 until 2017, and subtracted background. 
\cite{hebbar2019} reanalyzed the same data as \cite{vogt2018} but fit unbinned spectra, used time-dependent response files for the individual data sets, and carefully modeled the background in the complex environment around this feature. They also explored the sensitivity of their results to the correction for the time-dependent response and different assumed background models. \cite{hebbar2019} find that they can reject a simple blackbody model at the 99\% confidence level and explored more complex models such as a power-law plus a blackbody model and NS atmosphere models. They find their best fit with a C atmosphere model with a magnetic field of $B=10^{12}~\mathrm{G}$. They note that their best fitted thermal luminosity of $L=1.1\times10^{34}~\mathrm{ergs\, s^{-1}}$ and temperature of $T_\mathit{eff}=3.0\times10^6~K$ are higher than for most other NSs.

In this paper we present both a spectral and an imaging analysis of the X-ray properties of the putative CCO in E0102.  Our spectral analysis differs from that in \cite{vogt2018} in several significant ways but shares some of the methods adopted by \cite{hebbar2019}.  First, we generate response files for each observation that account for the changes in the effective area of the {\em Advanced CCD Imaging Spectrometer} (ACIS) as a function of time over the mission.  Given the faintness of the source, it is necessary to analyze data from multiple observations in order to accumulate sufficient counts for the analysis.  The data span a time range from 2003 to 2017, over which the ACIS effective area has changed dramatically at low energies. Second, we explicitly model the background spectrum while fitting the source spectrum.  The putative source is in a region of relatively high local background and the source flux is comparable to the surrounding background flux.  Therefore, in order to draw any meaningful conclusions about the source spectrum, it is necessary to carefully model the background.
Third, we preserve the spectral resolution of the data to maintain the sensitivity to line-like features by not binning the spectra.  
Fourth, we explore thermal spectral models to investigate if the spectral properties of this feature are consistent with thermal emission typical of the surrounding regions.
For the imaging analysis, we exploit the superb imaging capabilities of \chandra\/ to extract a radial distribution of the counts centered on the position of the claimed point source and compare that to the distribution expected for a point source in a region with background of comparable brightness to the source.

This paper is organized as follows.  In
\S \ref{sec:data.reduction} we describe the {\em Chandra} ACIS observations of E0102,
our detailed spectral analysis including our model for the background, and the generation of appropriate time-dependent response files that account for the changes in the effective area.
In \S\ref{sec:spectral.analysis}  we 
describe the results of our spectral fits with  blackbody models and thermal, non-equilibrium ionization models.
In Section \S\ref{sec:imganalysis}  we describe our
analysis of the imaging data comparing the distribution of the source counts to that expected for a point source.
In \S\ref{sec:discussion} and \S\ref{sec:conclusions} we
summarize our results and conclude.  Throughout this
paper we assume the distance to the SMC is 60.6 kpc \citep{hilditch2005}.
Error bars in plots and uncertainties on numerical values 
are quoted at the 1$\sigma$ confidence level unless otherwise stated.

\section{X-Ray Data and Reduction}\label{sec:data.reduction}
We analyze the observations used in \cite{vogt2018} for our spectral analysis, which are 25 ACIS-S3 on-axis observations (listed in Table~\ref{tbl:obslist}) in VFAINT mode, focal plane temperature variation less than $2^{\circ}$C, and no indication of background flares. We reprocessed each observation to generate new level=2 event lists, using \ciao 4.11 and CALDB 4.8.5 and the \ciao tool \texttt{chandra\_repro}\footnote{https://cxc.cfa.harvard.edu/ciao/ahelp/chandra\_repro.html} . The spectra were extracted for the source and background regions and the corresponding response files were created using \texttt{specextract}, and analyzed in \xspec version 12.11.0k \citep{arnaud1996}. In the image analysis, we used 10 observations (indicated by the footnote in Table~\ref{tbl:obslist}) which are used in the forward shock expansion measurement 
described in \citet{long2019} since these observations have been registered to each other. \marx 5.3 and \saotrace 2.0.4 were used for simulating point source and background events files in Section~\ref{sec:imganalysis}. We fit the unbinned spectra and use the C statistic \citep{cash1979} to avoid the bias introduced by the $\chi^2$ statistic in the case of low number of counts per bin \citep{kaastra2017}. 

We use the C statistic to determine the model parameters which provide the best fit and we calculate the value of the Pearson chi-square ($\chi^{2}_{P}$) for the model with these parameters held fixed to the values determined by the fit with the C statistic. We report both the value of $\chi^{2}_{P}$ and the value of the \texttt{goodness} command from \xspec as estimates of the goodness of fit.
The \texttt{goodness} command performs Monte Carlo simulations based on the model with the best fitted parameters and reports the fraction of simulated data sets that have a lower value of the fit statistic than the data
(\texttt{goodness} values close to 0.5 indicate a good fit, values close to 1.0 indicate a poor fit).
We fit all spectra simultaneously with the same model but each spectrum has response files appropriate for the instrument response given the date of the observation.
As we show in the following section \S\ref{sec:spectral.analysis}, it is crucial to use response files that are appropriate for the date of the observation.

\begin{table}[htbp]
\caption{Observations used in the spectral and imaging analysis. We used all 25 observations in \cite{vogt2018} in our spectral analysis, and 10 observations in our imaging analysis.}
\begin{center}
\begin{tabular}{l c c c}
\hline\hline
OBSID & Exposure & Roll angle & Start time\\
\hline
\phantom{0}3519 & \phantom{0}8.0 & 248 & 2003-02-01\\
\phantom{0}3520 & \phantom{0}7.6 & 248 & 2003-02-01\\
\phantom{0}3544 & \phantom{0}7.9 & \phantom{0}61 & 2003-08-10\\
\phantom{0}3545\footnote{Used in imaging analysis.\label{note.img}} & \phantom{0}7.9 & \phantom{0}64 & 2003-08-08\\
\phantom{0}5130 & 19.4 & 177 & 2004-04-09\\
\phantom{0}5131 & \phantom{0}8.0 & 181 & 2004-04-05\\
\phantom{0}6042 & 18.9 & 175 & 2005-04-12\\
\phantom{0}6043 & \phantom{0}7.9 & 169 & 2005-04-18\\
\phantom{0}6075 & \phantom{0}7.9 & 296 & 2004-12-18\\
\phantom{0}6758 & \phantom{0}8.1 & 198 & 2006-03-19\\
\phantom{0}6759 & 17.9 & 195 & 2006-03-21\\
\phantom{0}6765\footref{note.img} & \phantom{0}7.6 & 198 & 2006-03-19\\
\phantom{0}6766 & 19.7 & 126 & 2006-06-06\\
\phantom{0}8365 & 21.0 & 236 & 2007-02-11\\
\phantom{0}9694\footref{note.img} & 19.2 & 241 & 2008-02-07\\
10654 & \phantom{0}7.3 & 217 & 2009-03-01\\
10655 & \phantom{0}6.8 & 216 & 2009-03-01\\
10656 & \phantom{0}7.8 & 211 & 2009-03-06\\
11957\footref{note.img} & 18.4 & 283 & 2009-12-30\\
13093\footref{note.img} & 19.0 & 248 & 2011-02-01\\
14258\footref{note.img} & 19.0 & 270 & 2012-01-12\\
15467\footref{note.img} & 19.1 & 251 & 2013-01-28\\
16589\footref{note.img} & \phantom{0}9.6 & 190 & 2014-03-27\\
17380\footref{note.img}$^{,}$\footnote{Not used in spectral analysis.} & 17.7 & 228 & 2015-02-28\\
18418\footref{note.img} & 14.3 & 201 & 2016-03-15\\
19850 & 14.3 & 198 & 2017-03-19\\
\hline
\end{tabular}
\end{center}
\label{tbl:obslist}
\end{table}

\section{Spectral Analysis}\label{sec:spectral.analysis}
\begin{figure}[htb!]
  \centering
   \includegraphics[width=0.45\textwidth]
   {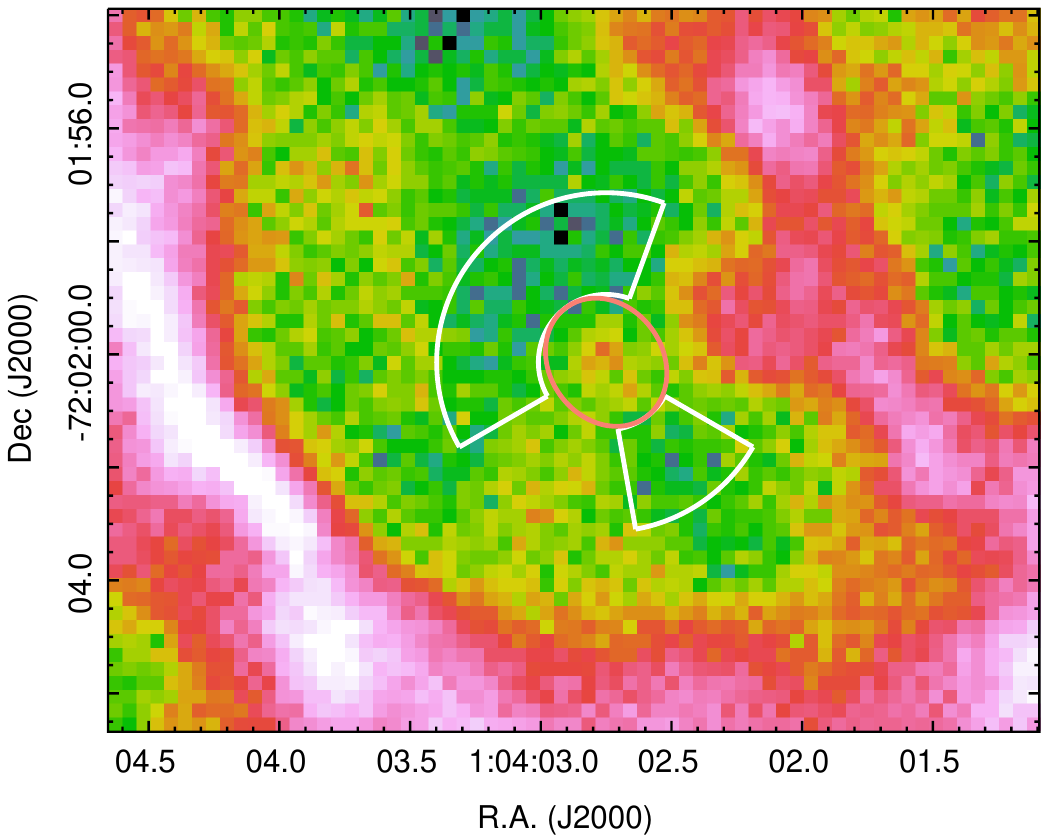}
   \caption{The source and near background region. The salmon ellipse is the source region, the white regions are near background regions. The image pixel size is half a sky pixel ($0.246\arcsec$), the energy band is in 0.35--4.0 keV.}
   \label{fig:srcbkg}
\end{figure}

\begin{figure}[htb!]
  \centering
   \includegraphics[width=0.45\textwidth]
   {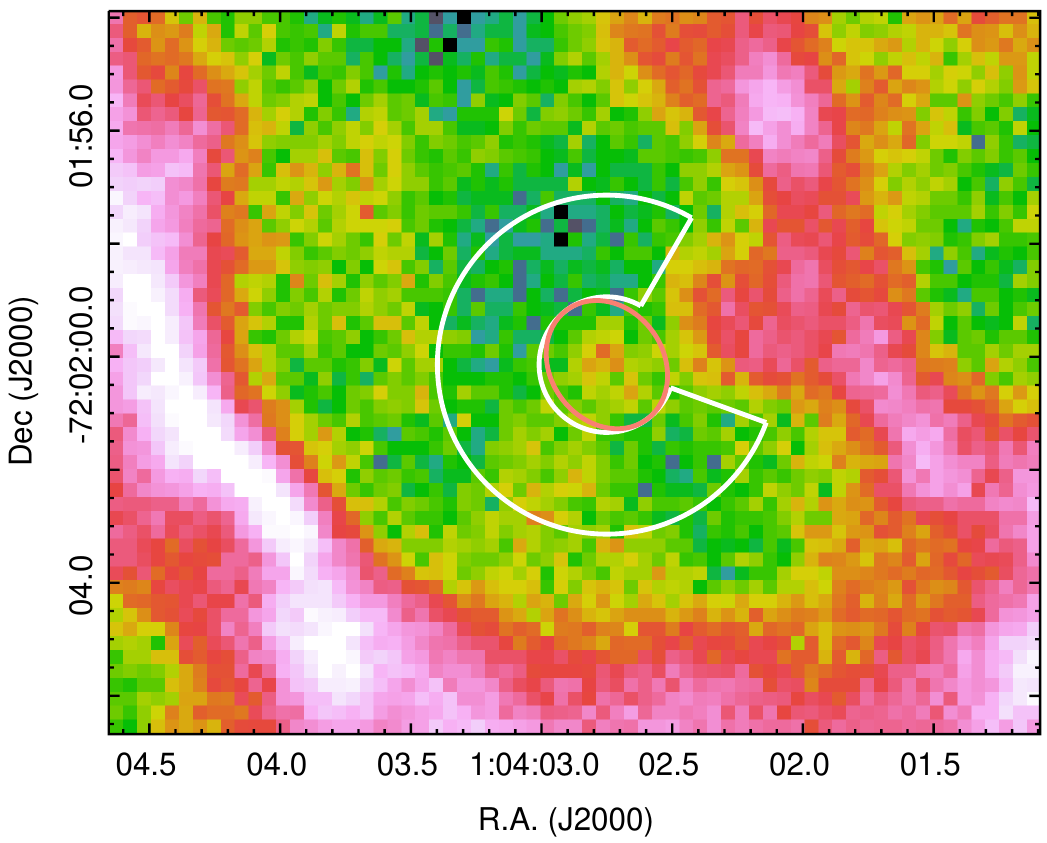}
   \caption{The source and near background region. The salmon ellipse is source region, the white region are near background regions. The image pixel size is half a sky pixel ($0.246\arcsec$), the energy band is in 0.35--4.0 keV.}
   \label{fig:srcbkg.alternative}
\end{figure}

We fit the source and background spectra simultaneously instead of subtracting background. For each observation, we have three spectra, one source spectrum and two background spectra from different background regions. The source spectrum is extracted from the elliptical region centered at $\alpha_{J2000}=01^{h}04^{m}02.75^{s}$, $\delta_{J2000}=-72^{\circ}02^{\prime}00.14^{\prime\prime}$, with a semi-major axis of $1.2^{\prime\prime}$ and a semi-minor axis of $1.0^{\prime\prime}$, identical to the region in \cite{vogt2018} as shown in Fig~\ref{fig:srcbkg}. One of the background spectra is extracted near the source region to model the emission from the region around the source, called the ``near background''.  We define two different extraction regions for the near background spectrum to allow an exploration of the sensitivity of our results to the selection of this background region, which is the dominant background component.  The first near background region is a two annular sector region around the source shown in Fig~\ref{fig:srcbkg} and the second near background region is the partial annular sector region shown in Fig~\ref{fig:srcbkg.alternative}.  The second near background region includes regions with higher surface brightness than the first near background region. We report results for the one sector and two sector background spectra and models in the tables that follow but we only show plots of spectral fits with the one sector background in the main body of the paper.  The plots of the spectral fits with the two sector background are similar to the fits with the one sector background and are included in Appendix~\ref{appdx:suppl.spectral.analysis} for completeness.

The other background spectrum is extracted off the remnant, in order to sample the sky and detector background. The sky and detector background region is a circular region $260^{\prime\prime}$ away from the center of the source region and off of the SNR. This component contributes little to the background below energies of 5.0~keV but is included for completeness.

\begin{figure}[htb!]
  \centering
   \includegraphics[width=0.45\textwidth]
   {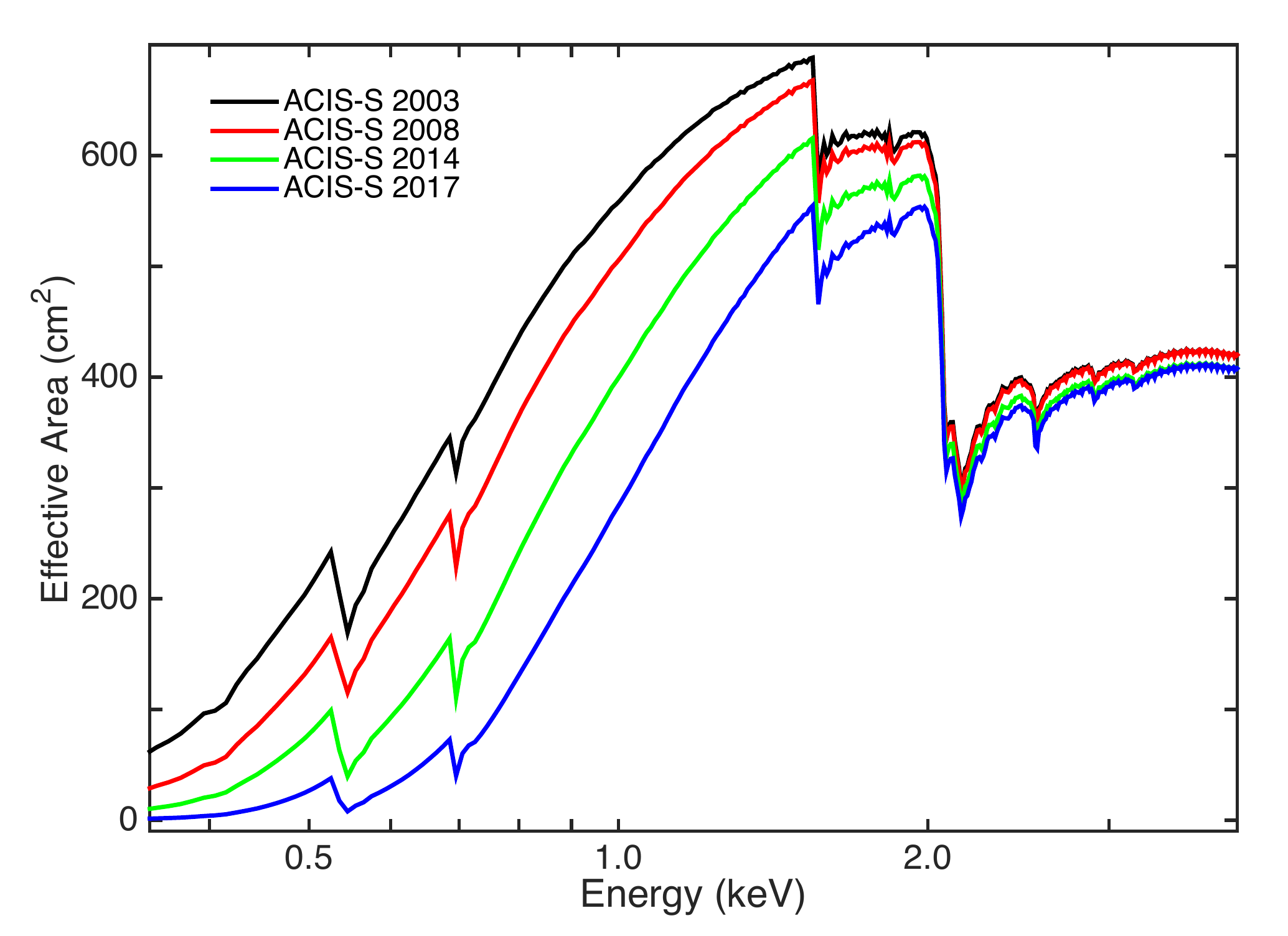}
   \caption{Effective area of ACIS-S in 2003, 2008, 2014, 2017 in the energy band 0.35--4.0 keV.  The effective area decreases with time below 2 keV, as a result of contamination accumulating on the optical blocking filter.}
   \label{fig:effarea}
\end{figure}

The ACIS effective area has changed significantly over the time period included in this analysis (2003-2017) due to the accumulation of a contamination layer on the optical blocking filter (OBF) in front of the CCDs \citep{plucinsky2018}.
Four examples of the effective area from 2003 to 2017 are shown in Fig~\ref{fig:effarea}. The effective area is significantly different between these four curves in the energy band 0.35--2 keV, which contains the majority of the  X-ray emission from E0102.  For example, the effective area at 654~eV, the energy of the bright \ion{O}{8}~Ly$\alpha$ line, decreases by about a factor of six over this time period.  The \ciao tools that generate the response products for spectral analysis account for the decreasing effective area with time by using the date of the observation to calculate the effective area. This makes it imperative that response files are generated for each observation individually as opposed to using a single response file for 2017 as was done in \citet{vogt2018}.
Thus, we fit the spectra for the 25 observations simultaneously  with appropriate time dependent response files applied to each observation.

\subsection{Near Background}\label{subsec.nearbkg}

We extract spectra from the two different near background regions, fit the spectra, and compare the results to characterize the dependence of the background model on the selected extraction region.
The first background region, as shown in Figure~\ref{fig:srcbkg}, consists of two annular sectors in order to include the diffuse emission around the source, but to exclude the
brighter regions to the west and southeast which appear to be associated with enhanced ejecta emission.
The second background region, as shown in Figure~\ref{fig:srcbkg.alternative}, includes the brighter
emission to the southeast.
The near background spectrum was fitted with a two \texttt{vnei} component model in \xspec. In this model, \texttt{wilm} abundances \citep{wilm2000} and the \texttt{vern} photoelectric constants \citep{verner1996} are used. The abundances of O, Ne, Mg, Si, S and the ionization timescale in the two \texttt{vnei} components are linked, and the other elements are set to typical SMC abundances (0.2). The temperatures and normalizations for the two components are allowed to vary. For the absorption we use the {\tt tbabs}
and {\tt tbvarabs} models, with the {\tt tbabs} component for the Galactic line of sight absorption, $N_{H},_{Galactic}=5.36 \times 10^{20} \mathrm{cm}^{-2}$ \citep{dickey1990}, and the {\tt tbvarabs} component for the SMC absorption with $N_{H},_{SMC}=5.76 \times 10^{20} \mathrm{cm}^{-2}$, determined from the XMM-Newton Reflection Gratings Spectrometer data presented in \cite{plucinsky2017}, with the SMC abundances specified in \cite{russell1992}.
Both absorption components were held fixed. The ``sky'' background component was fit with a {\tt constant*tbabs*tbabs*vnei} model in \xspec with the $N_{H}$ values set to the same values as for the near background model.  The ``detector'' background component was fit with a constant times a power law and multiple Gaussians for the instrumental lines similar to the approach in \cite{sharda2020}.  The sky and detector components were fit to the spectrum extracted from the background region off of the remnant to determine the best fitted parameters.  These parameters were then frozen when fitting 
the near background spectrum except for the two constant factors in each model.  The sky and detector components are a minor contribution to the background as shown below and only contribute significantly at energies above 5.0~keV.

\begin{figure}[htb!]
  \centering
   \includegraphics[width=0.45\textwidth]
   {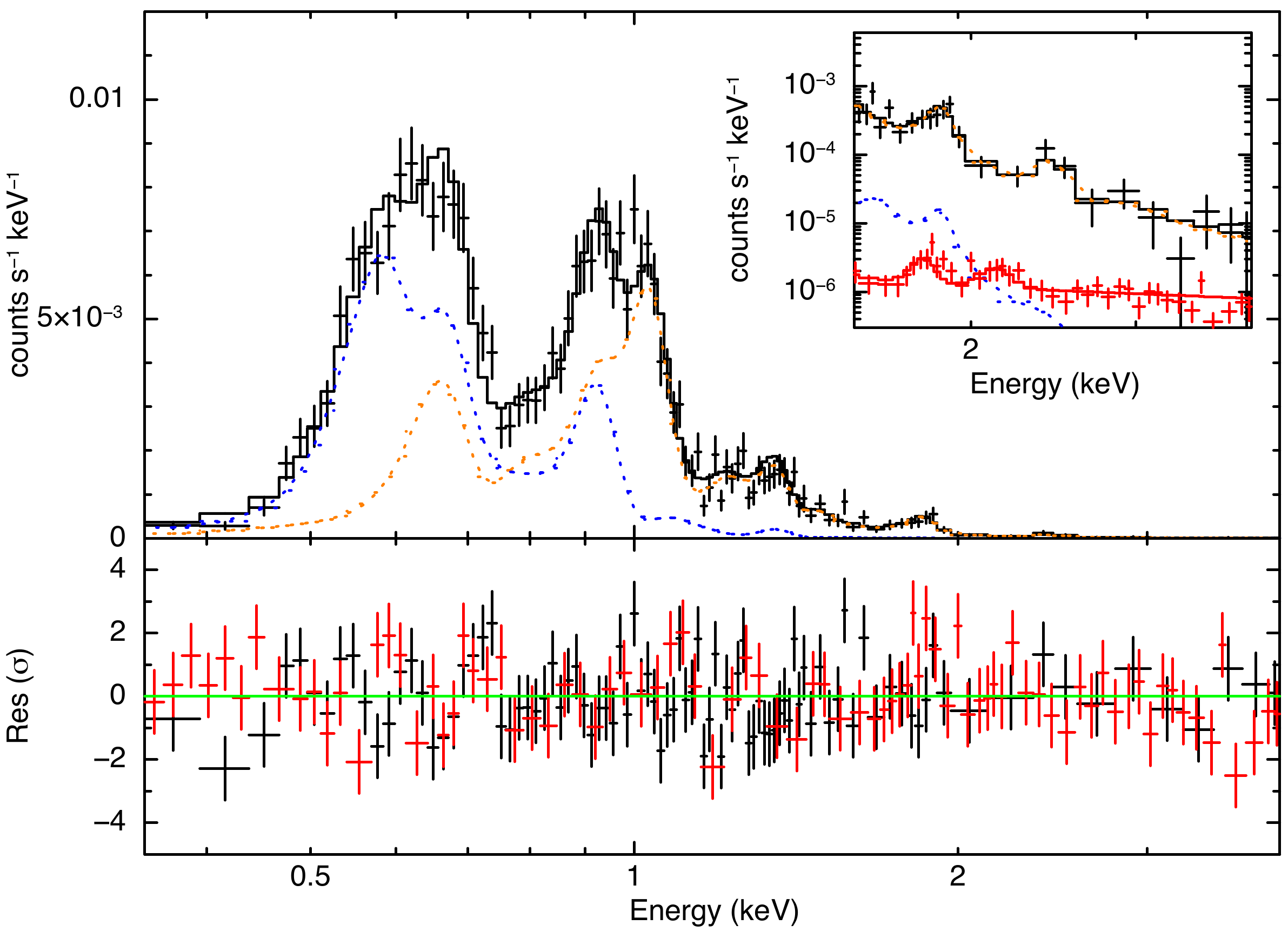}
   \caption{The near background spectrum from the two sector background region fit with the two \texttt{vnei} model. The black points and line are the near background data and model. The orange and blue dotted lines are the two \texttt{vnei}  components. The red points and line are the sky and detector background data and model. The inset shows the region from 1.5 to 4.0 keV. Note that the sky and detector background component is not visible in the main plot but is visible in the inset.  The black and red data points in the residual panel are the residuals for the near background and the sky and detector background models respectively.
   }
   \label{fig:nearback}
\end{figure}

\begin{figure}[htb!]
  \centering
   \includegraphics[width=0.45\textwidth]
   {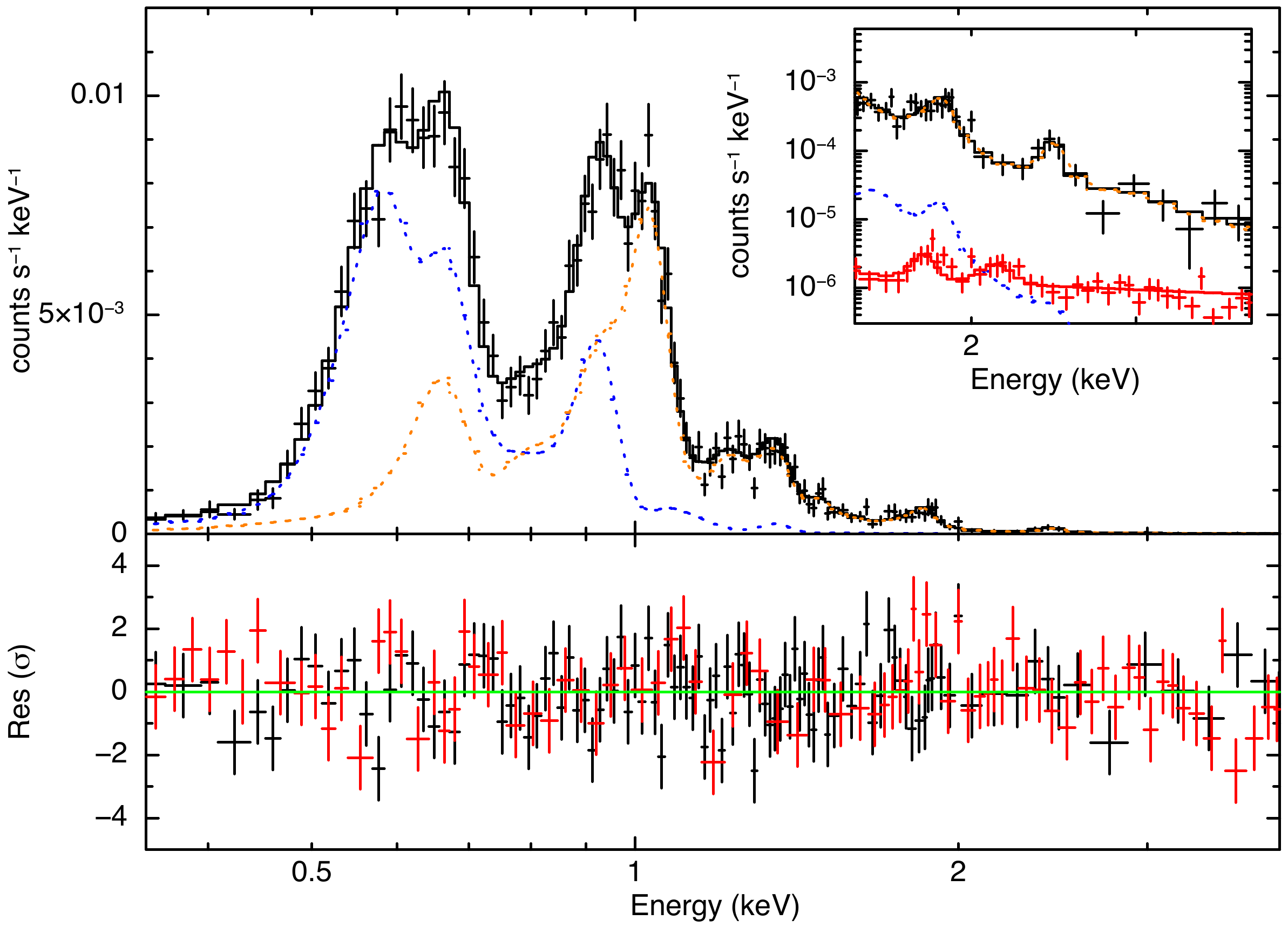}
   \caption{The near background spectrum from the one sector background region fit with the two \texttt{vnei} model.
   The black points and line are the near background data and model.
   The orange and blue dotted lines are the two \texttt{vnei} components. The red points and line are the sky and detector background data and model. The inset shows the region from 1.5 to 4.0 keV. Note that the sky and detector background component is not visible in the main plot but is visible in the inset.  The black and red data points in the residual panel are the residuals for the near background and the sky and detector background models respectively.
    }
   \label{fig:nearback.alternative}
\end{figure}

A single \texttt{vnei} component model does not provide an adequate fit to the near background data, hence we adopted a two component model. A single component model rarely provides an acceptable fit to a spectrum of the ejecta emission in E0102 since the O and Ne emission frequently require different values of the temperature and/or ionization timescale to achieve an acceptable fit.
The two component \texttt{vnei} model is intended to represent the complex structure of the shocked ejecta along this line of sight with different temperatures but with the same abundances and ionization time scales.
We seek the simplest model for the near background spectrum that provides an acceptable fit.
The fitted results are shown in Table~\ref{tbl:2.component.nearback} and the spectral fit in Figure~\ref{fig:nearback} for the two sector background extraction region and in Table~\ref{tbl:2.component.nearback} and Figure~\ref{fig:nearback.alternative} for the single sector background extraction region. The individual spectra are fitted simultaneously but grouped for display purposes only. It is clear from these spectra that the near background region has strong emission lines from O, Ne, and Mg which is similar to most of the bright ring in E0102.   The thermal emission modelled by the two \texttt{vnei} components dominates over the sky and detector background models in this bandpass. The detector background begins to dominate above 5.0~keV. Both spectra are fit well by the model but the two sector background spectrum is fit slightly better.

The fitted values for model of the two and one sector background regions are consistent within the $1\sigma$ uncertainties. The normalizations for the one sector near background model are higher than those of the two sector near background model as expected based on the images. The O, Ne, Mg, Si, and S abundances are all significantly enhanced with respect to typical SMC abundances indicating that the regions around the compact feature have a significant ejecta contribution. Enhanced abundances such as these are typical for regions with ejecta that have been heated by the reverse shock in E0102
\citep{alan2019}.
In the analysis that follows, we fit the source spectrum with both background models and demonstrate that the final results do not depend on the selection of background model.

\begin{table}[htbp]
\caption{Two-Component Spectral Model Parameters for the Near Background. The O, Ne, Mg, Si, S abundances and the ionization time scale are linked between the two \texttt{vnei} component.}
\begin{center}
\begin{tabular}{ l c c }
\hline\hline
Parameters & two sectors & one sector \\
\hline 
${kT_{e1}}\,\mathrm(keV)$ & $1.38^{+0.42}_{-0.25}$ & $1.37^{+0.28}_{-0.21}$\\
${kT_{e2}}\,\mathrm(keV)$ & $0.33^{+0.01}_{-0.03}$ & $0.31^{+0.01}_{-0.02}$\\
$n_\mathit{e}\,t,(10^{10}\,\mathrm{cm}^{-3}\mathrm{s})$ & $6.20^{+2.68}_{-1.80}$ & $7.08^{+1.85}_{-1.75}$\\
$Norm_{1},(10^{-6})$ & $2.47^{+0.74}_{-0.69}$ & $3.12^{+0.71}_{-0.62}$\\
$Norm_{2},(10^{-6})$ & $2.53^{+0.83}_{-0.76}$ & $3.26^{+0.82}_{-0.36}$\\
Oxygen & $ 1.86^{+0.31}_{-0.31}$ & $2.03^{+0.36}_{-0.28}$\\
Neon & $ 3.77^{+1.10}_{-0.71}$ & $4.15^{+0.76}_{-0.62}$\\
Magnesium & $ 2.10^{+0.70}_{-0.45}$ & $2.03^{+0.46}_{-0.34}$\\
Silicon & $1.51^{+0.54}_{-0.37}$  & $1.43^{+0.37}_{-0.29}$\\
Sulfur & $1.25^{+0.71}_{-0.51}$ & $1.48^{+0.56}_{-0.44}$\\
\textit{C-statistic}& 6830 & 7338\\
\textit{$\chi^2_{P}$} & 12676 & 12680\\
dof & 12438& 12438\\
goodness & 0.71& 0.76\\
\hline
\end{tabular}
\end{center}
\label{tbl:2.component.nearback}
\end{table}

\subsection{Source Spectrum}
\begin{figure}[htb!]
  \centering
   \includegraphics[width=0.45\textwidth]
   {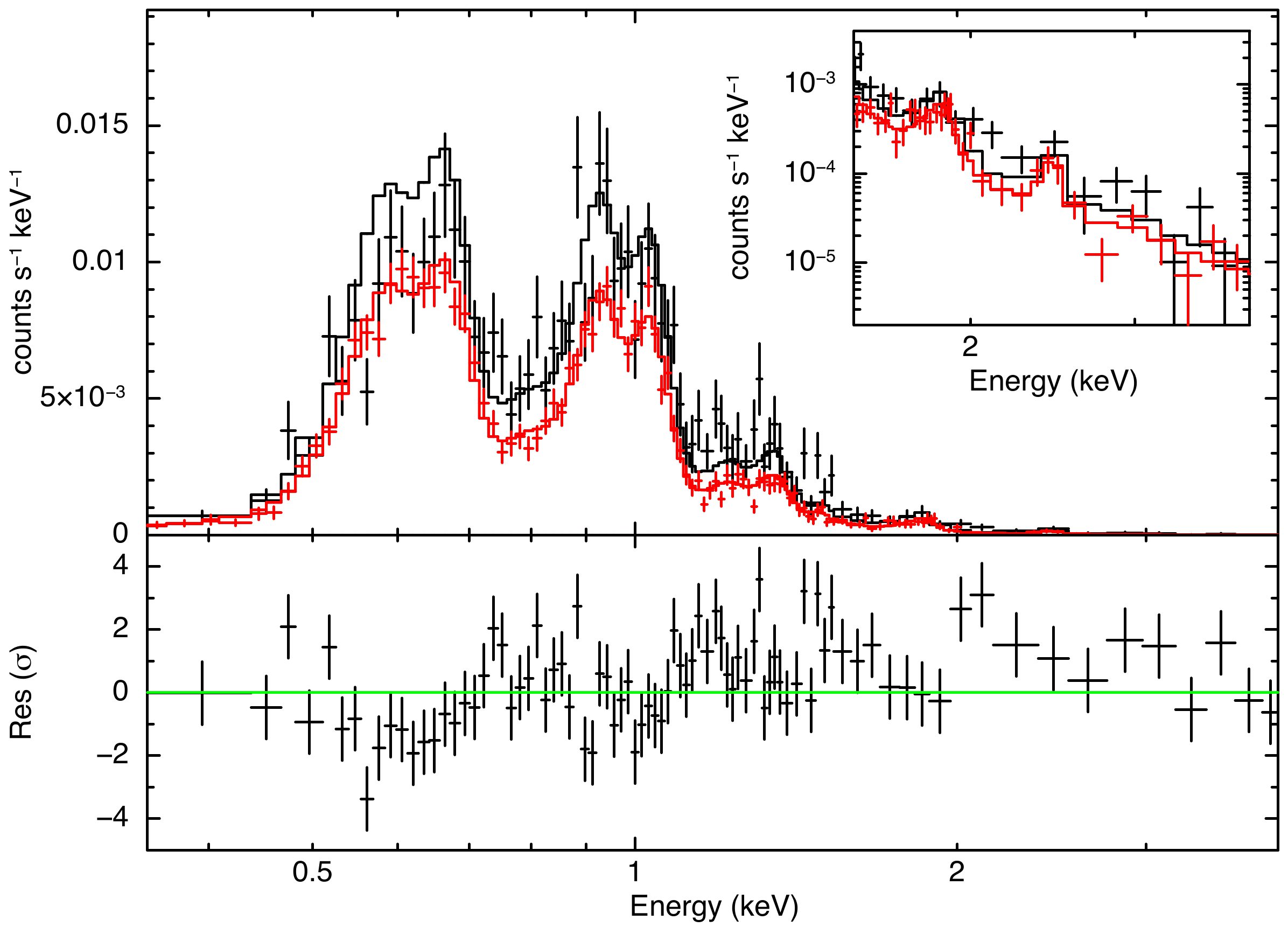}
   \caption{Source spectrum fit with the one sector near background model.
   The black points are the source data and the black line is the fitted near background model with only the normalization free. The red points and line are the near background data and model respectively. 
   }
   \label{fig:scale.alternative.nearback}
\end{figure}

We first test the hypothesis that the source and the near background spectrum have the same intrinsic shape but differ only in intensity.  This might be the case if these two regions were part of the same, larger structure in the remnant but with different intensities. We fit the source spectrum with the near background spectrum model, allowing only a global normalization to vary.  The result is shown for the one sector near background spectrum in Figure~\ref{fig:scale.alternative.nearback} and the two sector near background spectrum
in Figure~\ref{fig:scale.nearback}. 
It is clear from these plots that the spectrum from the compact feature is
only slightly higher in intensity than the near background spectrum.  This demonstrates how important it is to properly model the background as opposed to subtracting the background.  It is also clear that the shape of the source spectrum is similar to the shape of the near background spectrum, which is not surprising if the near background contributes the majority of the counts in the source spectrum.
As expected the model background level is higher for the one sector background model since it includes the region of higher surface brightness in the southeast as shown in Figure~ \ref{fig:srcbkg.alternative}.
Although the shape of the near background spectrum is similar to the source spectrum, there are clear differences between the two.
The near background model can not fit the source spectrum at O, Mg, and above 2 keV as can be seen in the residuals for the fit. 
The value of the C statistic is 10251, the value of the $\chi^{2}_{P}$ is 21092, and the number of degrees of freedom is 18671. The \texttt{goodness} command indicates that more than 98\% of the simulated data sets have a test statistic value lower than this value.
We conclude that the source spectrum has a different shape than that of the near background spectrum and explore different models for the source spectrum in the following sections.

\subsubsection{Source Spectrum fit with blackbody Model}\label{subsubsec.bb}
\begin{figure}[htb!]
  \centering
   \includegraphics[width=0.45\textwidth]
   {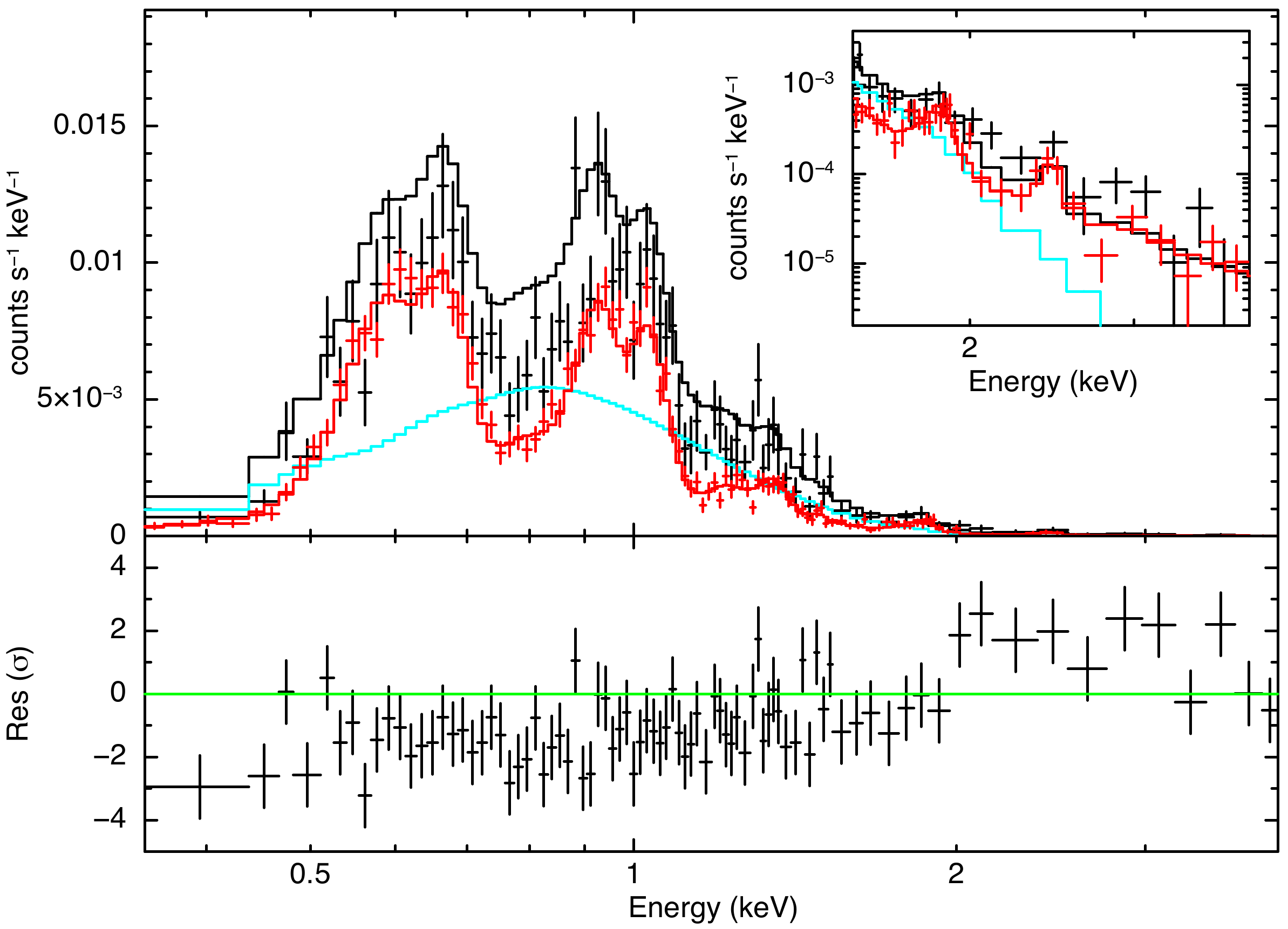}
   \caption{Source spectrum fit with a blackbody model and the one sector near background model. The blackbody parameters are fixed to the values in \cite{vogt2018} and the normalization of the near background component (linked for the source and near background models) is allowed to vary.  The black points and line are the source data and model. The red points and line are the near background data and model. The cyan line is the blackbody component in the source model. The blackbody model overpredicts the source spectrum in the 700-900~eV range and underpredicts above 2.0~keV.}
   \label{fig:srcbbnewnb}
\end{figure}

\begin{figure}[htb!]
  \centering
   \includegraphics[width=0.45\textwidth]
   {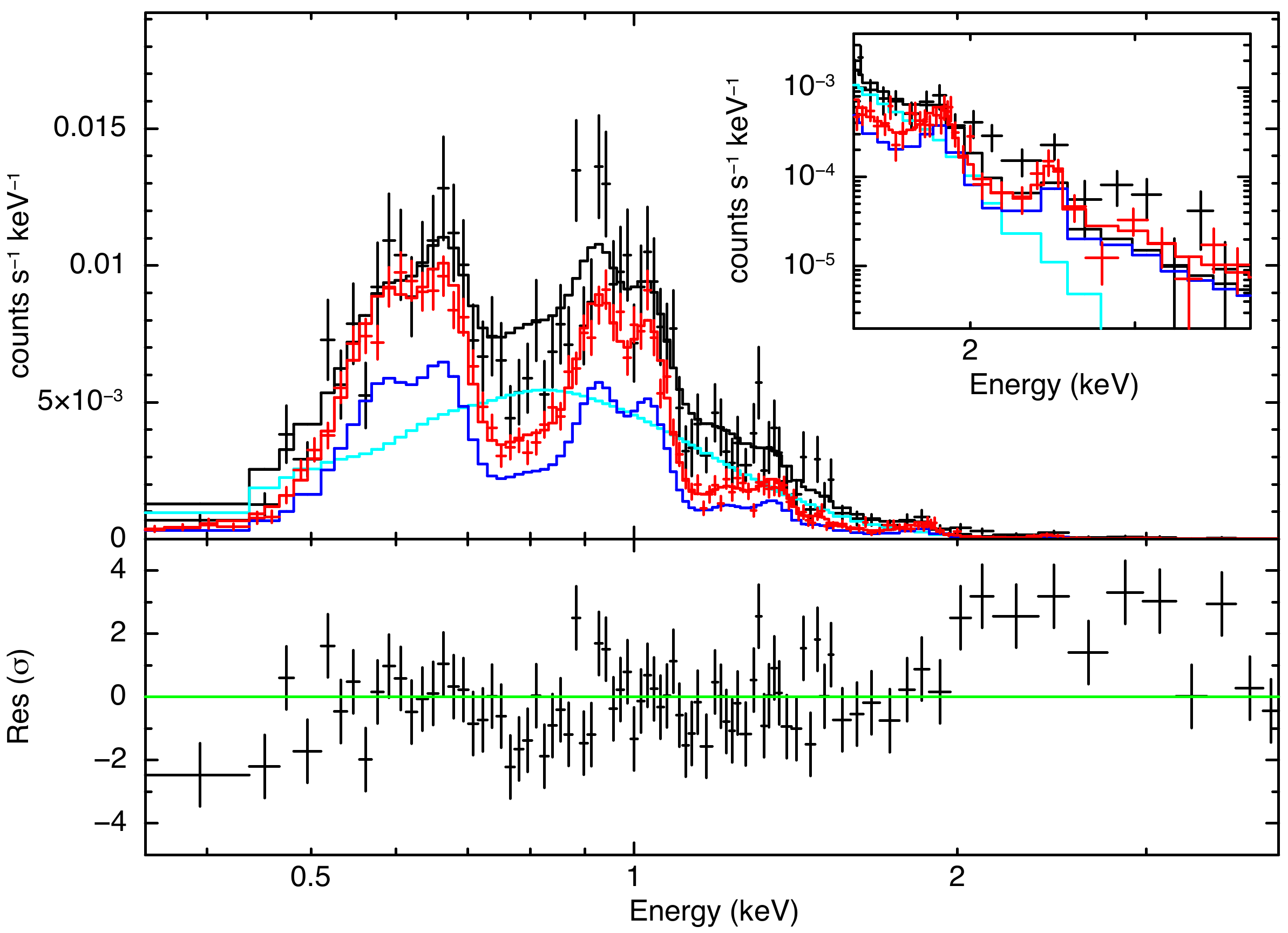}
   \caption{Source spectrum fit with a blackbody model and the one sector near background model. The blackbody parameters are fixed to the values in \cite{vogt2018} and the normalizations of the near background component in the  source and near background models are allowed to vary independently.
   The black points and line are the source data and model. The red points and line are the near background data and model. The cyan line is the blackbody component in the source model. The blue line is the near background component in the source model.}
   \label{fig:srcbbfnbnewnb}
\end{figure}

\begin{figure}[htb!]
  \centering
   \includegraphics[width=0.45\textwidth]
   {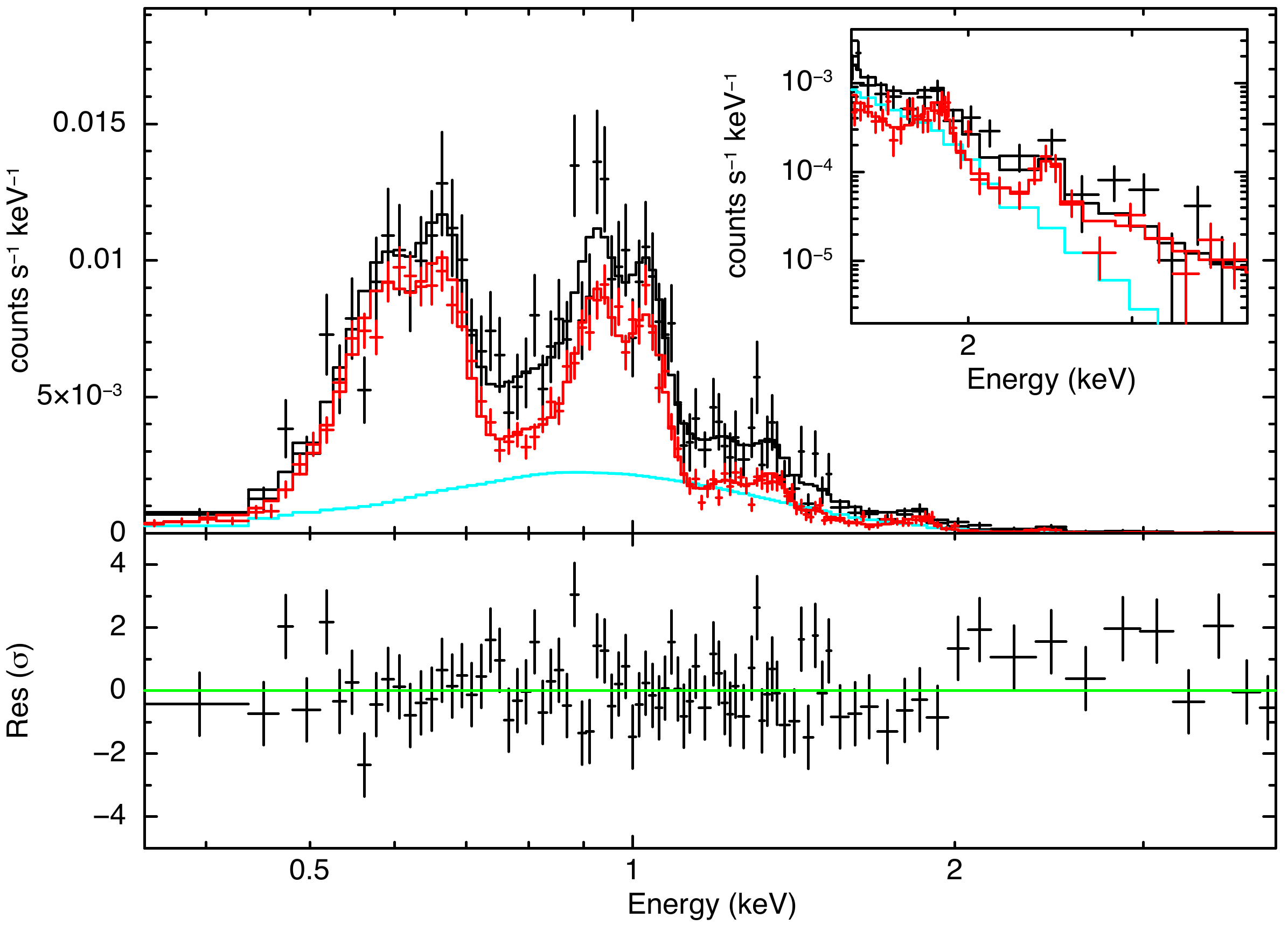} 
    \caption{Source spectrum fit with a blackbody model and the one sector near background model. The temperature and normalization of the blackbody component  and the normalization of the near background
    component 
    (linked for the source and near background models) are free in the fit. The black points and line are the source data and model. The red points are the near background data and model. The cyan line is the blackbody component in the source model.}
   
   \label{fig:srcbbfpnewnb}
\end{figure}

\begin{figure}[htb!]
  \centering
   \includegraphics[width=0.45\textwidth]
   {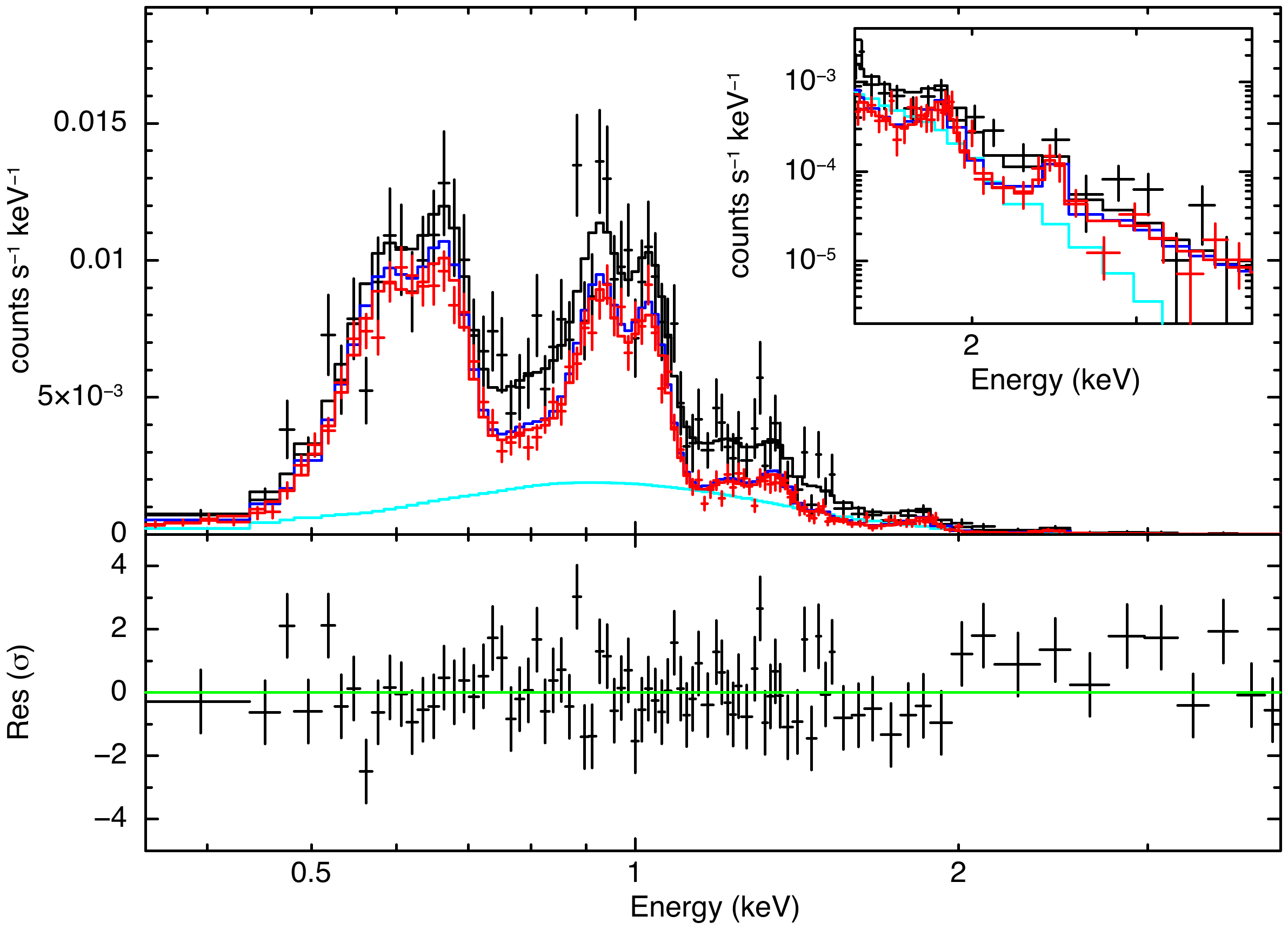}
   \caption{Source spectrum fit with a blackbody model and the one sector near background model. The temperature and normalization of the blackbody component  are free in the fit and the normalizations of the near background component in the  source and near background models are allowed to vary independently. The black points and line are the source data and model. The red points and line are the near background data and model. The cyan line is the blackbody component in the source model. The blue line is the near background component in the source model.}
   \label{fig:srcfbbfnbnewnb}
\end{figure}

We fit the source spectrum with a blackbody model and our near background models.
We used the same two component absorption model for the blackbody spectrum as we used for the near background spectrum.
We first fit with the parameters for the blackbody model specified in \citet{vogt2018} and the near background model normalization free to vary, with the normalizations for the near background component in the source and near background models linked together.  We then repeat this fit allowing the near background model normalization for the source and near background models to vary
independently of each other,
allowing for the possibility that the near background normalization is different for the source region. We then explore allowing the temperature and normalization of the blackbody model to vary from the values in \citet{vogt2018} allowing for the possibility that the best fitted values are different in our analysis due to our different methods. For this fit, the normalizations for the near background component in the source and near background models are linked together.
Finally we repeat this fit with the normalization for the near background component in the source model free to vary independently of the normalization in the near background model.
We adopt this approach of multiple fits to two different background spectra extracted from two different regions with different assumptions for the near background model to examine the sensitivity of our final result on the background model.
The results are  summarized in Table~\ref{tab:srcbb} and the spectral fits  are shown in Figures~\ref{fig:srcbbnewnb}--\ref{fig:srcfbbfnbnewnb} for the one sector background and in
Figures~\ref{fig:srcbb}--\ref{fig:srcfbbfnb} for the two sector background.

\begin{table*}[htb!]
\caption{Source spectrum fit with a blackbody model and our near background models. The temperature and normalization of blackbody component are set to the Vogt values in two fits and allowed to vary in the other fits.}
\begin{center}
\label{tab:srcbb}
\begin{tabular}{l c c c c}
\hline\hline
&&two sectors near background&&\\
\hline
model & fixed T and norm & fixed T and norm & free T and norm  & free T and norm\\
  & & free near back scale &  & free near back scale\\
\hline 
temperature (keV) & 0.19 & 0.19 & $0.23^{+0.02}_{-0.01}$  & $0.26^{+0.03}_{-0.02}$ \\
normalization ($10^{-7}$) & 3.399  & 3.399 & $1.81^{+0.13}_{-0.12}$  &$1.09^{+0.22}_{-0.21}$\\
near background in source & $0.96^{+0.01}_{-0.01}$ & $0.74^{+0.03}_{-0.03}$ & $1.01^{+0.02}_{-0.01}$ & $1.27^{+0.08}_{-0.07}$ \\
near background scale & $0.96^{+0.01}_{-0.01}$ & $1.00^{+0.02}_{-0.02}$ & $1.01^{+0.02}_{-0.01}$  & $1.00^{+0.02}_{-0.02}$ \\
$\mathrm{Flux}_{(0.35-4.0 keV)}$($10^{-14} \mathrm{erg\,cm^{-2}\,s^{-1}}$)& 1.72 & 1.72 & $1.05^{+0.06}_{-0.06}$ & $0.68^{+0.10}_{-0.10}$\\
$\mathrm{L}_{(0.35-4.0 keV)}$($10^{33} \mathrm{erg\,s^{-1}}$)& 10.62 & 10.62 & $6.00^{+0.38}_{-0.39}$ & $3.71^{+0.71}_{-0.67}$\\
\textit{C-statistic}& 9805 & 9759 & 9690  & 9678 \\
$\chi^{2}_{P}$ & 23631  & 25671 & 22137 & 20618\\
dof & 18672 & 18671 & 18670  & 18669\\
goodness & 1.00 & 1.00 & 0.99 & 0.93\\
\hline
&&one sector near background&&\\
\hline
model & fixed T and norm  & fixed T and norm & free T and norm & free T and norm\\
  & & free near back scale & & free near back scale\\
\hline
temperature (keV) & 0.19  & 0.19 & $0.24^{+0.01}_{-0.01}$  & $0.25^{+0.02}_{-0.02}$ \\
normalization ($10^{-7}$) & 3.399  & 3.399 & $1.35^{+0.13}_{-0.12}$  &$1.16^{+0.25}_{-0.23}$\\
near background in source & $0.96^{+0.01}_{-0.01}$ & $0.64^{+0.03}_{-0.03}$ & $1.00^{+0.01}_{-0.01}$ & $1.06^{+0.06}_{-0.06}$ \\
near background scale & $0.96^{+0.01}_{-0.01}$ & $1.00^{+0.01}_{-0.01}$ & $1.00^{+0.01}_{-0.01}$ & $1.00^{+0.01}_{-0.01}$ \\
$\mathrm{Flux}_{(0.35-4.0 keV)}$($10^{-14} \mathrm{erg\,cm^{-2}\,s^{-1}}$)& 1.72 & 1.72 & $0.80^{+0.06}_{-0.06}$ & $0.70^{+0.12}_{-0.12}$ \\
$\mathrm{L}_{(0.35-4.0 keV)}$($10^{33} \mathrm{erg\,s^{-1}}$)& 10.62 & 10.62 & $4.53^{+0.39}_{-0.38}$ & $3.90^{+0.78}_{-0.74}$ \\
\textit{C-statistic} & 10363 & 10246 & 10181 & 10181\\
$\chi^{2}_{P}$ & 21834 & 24584 & 21026  & 20661\\
dof & 18672 & 18671 & 18670  & 18669\\
goodness & 0.99& 1.00 & 0.98 & 0.96\\
\hline
\end{tabular}
\end{center}
\end{table*}
The source spectrum fit with the
blackbody model with the temperature and normalization fixed at the values in \citet{vogt2018} are shown in Figures~\ref{fig:srcbbnewnb} and ~\ref{fig:srcbb} for the one sector and two sector near background models respectively, with the normalizations for the near background component linked together for the source and near background spectra.
The blackbody model overestimates the source spectrum at energies between 700 and 900~eV and underestimates the spectrum at energies above 2.0~keV for both near background models as
indicated by the residuals for the fit. These discrepancies highlight the importance of retaining the full spectral resolution of the detector when fitting spectral models.  The fitted values and fit statistics are shown in the second column of Table~\ref{tab:srcbb}.
The \texttt{goodness} value is 1.0 indicating an unacceptable fit. We repeat these fits with the normalization for the near background component in the source model allowed to vary independently of the normalization in the near background model. The spectral fits are shown in Figures~\ref{fig:srcbbfnbnewnb} and~\ref{fig:srcbbfnb} and
the results are listed in the third column of Table~\ref{tab:srcbb}. The normalization for the near background in the source model decreases to 0.74 and 0.64 from the previous value of 0.96 for the two sector and one sector near background models respectively but the \texttt{goodness} values of 1.0 indicate that both fits are unacceptable.
We conclude from these fits that simply rescaling the near background model for the source spectra without adjusting its shape will not provide an adequate fit if the blackbody parameters specified in \citet{vogt2018} are assumed for the source.

We next explore if allowing the blackbody parameters and the normalization to vary can produce acceptable fits. We first allow the temperature and normalization of the blackbody component to vary with the normalizations for the near background component linked together for the source and near background spectra. The spectral fits are shown in Figures~\ref{fig:srcbbfpnewnb} and~\ref{fig:srcbbfp} and
the results are listed in the fourth column of Table~\ref{tab:srcbb}. 
The blackbody temperature increases and the normalization decreases by a factor of 2-2.5 while the normalization for the near background component increases to 1.00 and 1.01.
The fit statistics improve significantly but the fits are still unacceptable.
The \texttt{goodness} values improve marginally to 0.99 and 0.98 for the two sector and one sector near background spectra respectively.
The residuals in Figures~\ref{fig:srcbbfpnewnb} and~\ref{fig:srcbbfp} 
show that the fit has improved in the 700-900~eV range but the model is still underpredicting at energies above 2.0~keV.

Finally, we consider the case in which the blackbody parameters are allowed to vary and the normalization of the near background model in the source model is allowed to vary independently of the normalization in the near background model.  The results in the fifth column of Table~\ref{tab:srcbb} show that the blackbody temperature increases to 0.26~keV and 0.25~keV and the normalization decreases by a factor of three compared to the \citet{vogt2018} value. The normalization for the near background in the source model increases to 1.27 and 1.06 for the two sector and one sector near background spectra respectively.  The fit and residuals are shown in Figures~\ref{fig:srcfbbfnbnewnb} and \ref{fig:srcfbbfnb}.  The blackbody model is harder and has its normalization reduced by a factor of three while the near background model normalization has increased  to make up the difference. Both fits can be rejected as the \texttt{goodness} values are 0.93 and 0.96.
We conclude that a single blackbody model spectrum can not provide an acceptable fit even if the temperature and normalization of the blackbody is allowed to vary and the normalization of the near background in the source model is allowed to vary independently of the normalization in the near background model.
This is consistent with the  \cite{hebbar2019} result in which they added a power-law component to their blackbody model to account for the emission above 2.0~keV in order to achieve an acceptable fit
(see discussion in \S\ref{sec:discussion}).

\subsubsection{Source Spectrum fit with a Non-Equilibrium Ionization Plasma Model}\label{subsubsec.vnei}
We now explore a thermal, non-equilibrium ionization (NEI) model with variable abundances for O, Ne, Mg, Si, S, and Fe for the source (the {\tt vnei} model in \xspec ).
We use the same two component absorption model for the {\tt vnei} model  spectrum as we used for the near background spectrum.
This model considers the possibility that  the source is similar to other regions in the SNR which are dominated by ejecta heated by the reverse shock.  Detailed analyses of the E0102 spectra have been presented in \citet{rasmussen2001}, \citet{sasaki2001}, \citet{flanagan04} \citet{plucinsky2017}, and \citet{alan2019}.

\begin{figure}[htb!]
  \centering
   \includegraphics[width=0.45\textwidth]
   {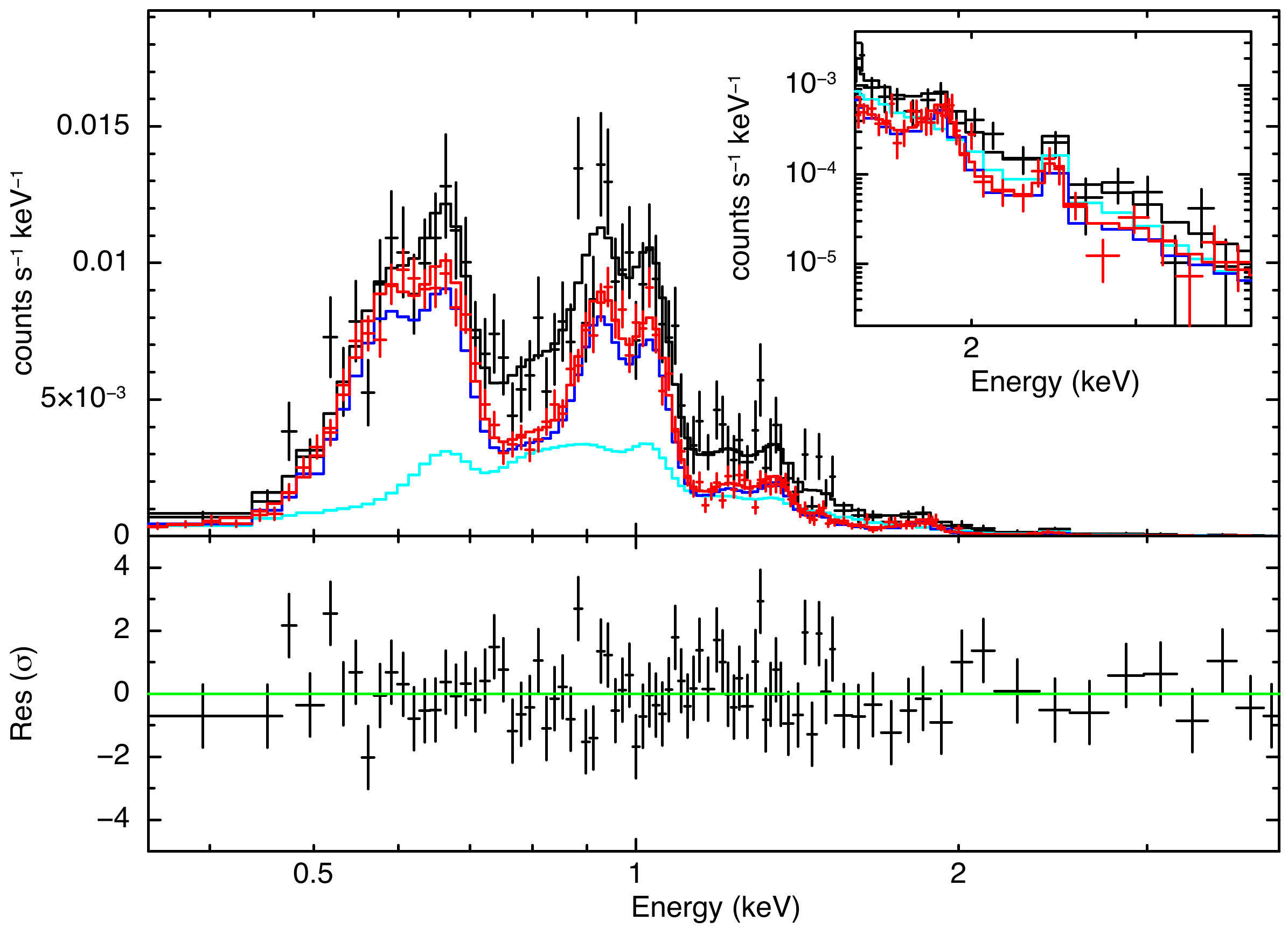}
   \caption{Source spectrum fit with a \texttt{vnei} model
   ($n_\mathit{e}\,t \sim3.3\times10^{11}\,\mathrm{cm}^{-3}\mathrm{s}$)
   and the one sector near background model. The black points and line are the source data and model. The red points and line are the near background data and model. The cyan line is the \texttt{vnei} component in the source model.
   The \texttt{vnei} model parameters are listed in Table~\ref{tbl:source.vnei}.}
   \label{fig:srcvnei.free.fe}
\end{figure}

There are two classes of {\tt vnei} models that provide comparable, acceptable fits to the data. One class has a temperature of 0.79-0.83 keV with an ionization timescale of $n_\mathit{e}\,t \sim3.3\times10^{11}\,\mathrm{cm}^{-3}\mathrm{s}$ and the second class has a temperature of 0.91-0.97 keV with an ionization timescale of $n_\mathit{e}\,t \sim7.0\times10^{10}\,\mathrm{cm}^{-3}\mathrm{s}$.
We will discuss the class of models with an ionization timescale of
$n_\mathit{e}\,t \sim3.3\times10^{11}\,\mathrm{cm}^{-3}\mathrm{s}$
first. The fit results are listed in Table~\ref{tbl:source.vnei} and 
the spectra and model fits are displayed in Figures~\ref{fig:srcvnei.free.fe} and~\ref{fig:srcvnei.free.fe.2}. 
As we have done throughout this paper, we include results for both the two sector and one sector background spectra.   The fit statistics listed in Table~\ref{tbl:source.vnei}  indicate that these fits are formally acceptable as the values of the $\chi^{2}_{P}$ (18670-18705) are quite close to the degrees of freedom (18662) and the \texttt{goodness} values range from 0.56 to 0.60. The O, Ne, and S abundances are marginally enhanced with respect to SMC interstellar medium (ISM) abundances however the uncertainties are large enough that these values are consistent with SMC ISM abundances within a $1\sigma$ uncertainty. The Si and Fe abundances are zero or close to zero.  It is not surprising that the fitted values of the abundances are not well-constrained given that 
the near background spectrum dominates the source
spectrum in intensity and the near background spectrum clearly shows line-like emission from O, Ne, Mg, Si, and S.
It should be noted that the abundances in Table~\ref{tbl:source.vnei} are lower than the abundances in 
Table~\ref{tbl:2.component.nearback}.
However, the fitted abundances in the source and near background models are correlated and any reduction in a fitted abundance in the near background model would result in an increase in the fitted abundance in the source model.
The ionization timescale is not constrained for high values as the best fitted value approaches the value for which
collisional ionization equilibrium applies and the limited statistics in the source spectrum do not allow a constraint on this value.  The derived values for the two sector and one sector background models are all consistent with each other within the $1 \sigma$ uncertainties. Therefore, this thermal model does provide an acceptable fit to the source spectrum and there is no evidence that the result depends on which background region is selected for the analysis.

\begin{figure}[htb!]
  \centering
   \includegraphics[width=0.45\textwidth]
   {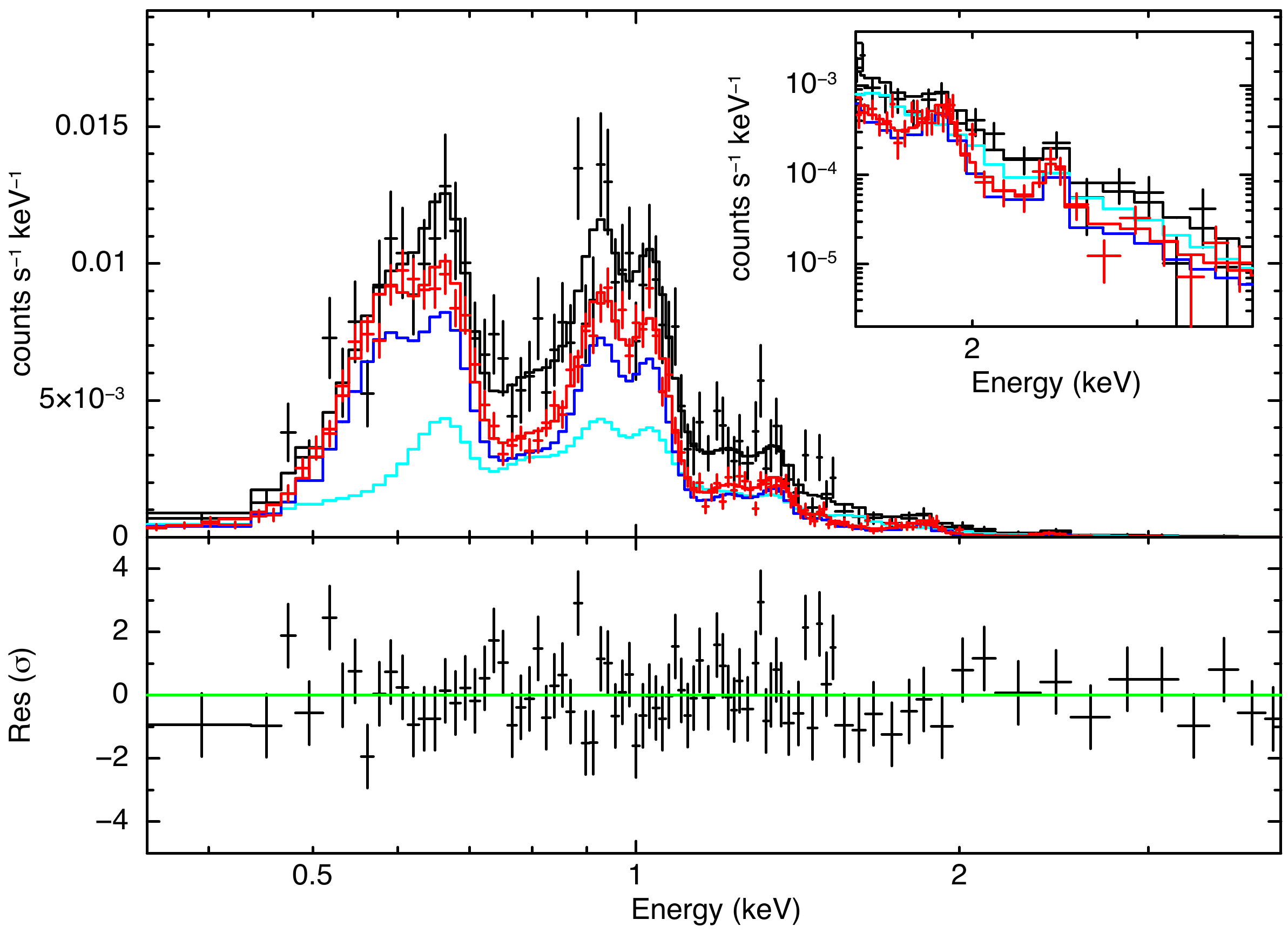}
   \caption{Source spectrum fit with a \texttt{vnei} model
    ($n_\mathit{e}\,t \sim7.0\times10^{10}\,\mathrm{cm}^{-3}\mathrm{s}$)
   and the one sector near background model. The black points and line are the source data and model. The red points and line are the near background data and model. The cyan line is the \texttt{vnei} component in the source model.
   The \texttt{vnei} model parameters are listed in \ref{tbl:source.vnei.nosi.nofe}.}
   \label{fig:srcvnei.nofe}
\end{figure}

Given the low values of the Si and Fe abundances in the previous fit, we fit the spectra with a simplified model in which the Si and Fe abundances were frozen at 0.0 and the S abundance was frozen at 0.2 and only the O, Ne, and Mg abundances were allowed to vary.  These are the second class of fits with a significantly lower value of the ionization timescale of 
$n_\mathit{e}\,t \sim7.0\times10^{10}\,\mathrm{cm}^{-3}\mathrm{s}$.  The fit results are shown in Table~\ref{tbl:source.vnei.nosi.nofe} and the spectra and models are shown in Figures~\ref{fig:srcvnei.nofe} and~\ref{fig:srcvnei.nofe.2} .  The O, Ne, and Mg abundances are now much lower and consistent with SMC ISM values. Although the model spectra still show clear line emission from O, Ne, and Mg as the emissivity of these lines is larger at the higher value of the temperature and lower value of the ionization timescale than in the previous fit. 
The abundances in Table~\ref{tbl:source.vnei.nosi.nofe} are now much lower than the abundances in Table~\ref{tbl:2.component.nearback},
but as mentioned the previous paragraph, the correlation between the abundances in the source and near background models also exists in these fits. The fit statistics  indicate that these fits are formally acceptable as the values of the $\chi^{2}_{P}$ (18486-18575) are quite close to the degrees of freedom (18665) and the \texttt{goodness} values range from 0.46 to 0.49.  The normalization for the near background component in the source model is significantly lower (0.81-0.98) and this has been compensated for by a higher normalization for the source spectrum. The derived values for the two sector and one sector background models are all consistent with each other within the $1 \sigma$ uncertainties. Therefore, this thermal model also provides an acceptable fit to the source spectrum and there is no evidence that the result depends on which background region is selected for the analysis.

Given the limited statistics in the source spectrum it is not possible to distinguish between these two classes of thermal models.  The fits with an ionization timescale of $n_\mathit{e}\,t \sim7.0\times10^{10}\,\mathrm{cm}^{-3}\mathrm{s}$ have slightly lower values of the fit statistics but the fit statistics do not differ by a large enough amount that the models with an ionization timescale of  $n_\mathit{e}\,t \sim3.3\times10^{11}\,\mathrm{cm}^{-3}\mathrm{s}$ can be ruled out.  In the first class of models the results are consistent with a plasma that has enhanced abundances of O and Ne and has been shocked for a time comparable with the age of the remnant, while in the second class of models the results are consistent with a plasma with SMC ISM abundances that has 
been shocked more recently.
These fits show that a thermal, NEI model can produce an acceptable fit to the source 
spectrum.

\begin{table}[htbp]
\caption{Source spectrum fit with the \texttt{vnei} model. 
The best fitted parameter values with the $1 \sigma$ uncertainties are listed for the two sector and one sector background models.
}
\begin{center}
\begin{tabular}{ l c c }
\hline\hline
Parameters & two sectors & one sector \\
\hline 
${kT_{e}}\,\mathrm(keV)$ & $0.83^{+0.17}_{-0.09}$ & $0.79^{+0.21}_{-0.08}$\\
$n_\mathit{e}\,t_{1},(10^{11}\,\mathrm{cm}^{-3}\mathrm{s})$ &   $3.34^{\dagger}_{-2.59}$ & $3.33^{\dagger}_{-2.59}$\\
$Norm,(10^{-5})$ &  $1.57^{+0.41}_{-0.38}$ &  $1.71^{+0.42}_{-0.36}$\\
Oxygen &  $ 1.53^{+1.47}_{-0.95}$  &  $1.33^{+1.04}_{-0.91}$\\
Neon &  $ 0.77^{+1.97}_{-0.51}$ &  $0.47^{+1.33}_{-0.25}$ \\
Magnesium &   $ 0.34^{+0.66}_{-0.21}$ &   $0.29^{+0.55}_{-0.17}$\\
Silicon &  $0.00^{+0.09}_{-0.00}$ &   $0.00^{+0.08}_{-0.00}$\\
Sulfur &  $1.04^{+0.88}_{-0.57}$ & $0.94^{+0.76}_{-0.58}$\\
Iron &   $ 0.08^{+0.06}_{-0.04}$ & $0.07^{+0.03}_{-0.03}$\\
near background in source &  $1.04^{+0.09}_{-0.09}$ & $0.90^{+0.08}_{-0.07}$\\
near background scale &  $1.00^{+0.01}_{-0.01}$ & $1.00^{+0.01}_{-0.01}$\\
\textit{C-statistic}&  9665 &  10170\\
\textit{$\chi^{2}_{P}$} &  18670 &  18705 \\
dof &  18662  & 18662\\
goodness & 0.56 & 0.60\\
\hline
\end{tabular}
\begin{tablenotes}
        \footnotesize
        \tablenotetext{\dagger} {The upper limit can not be constrained by the model}
\end{tablenotes}
\end{center}
\label{tbl:source.vnei}
\end{table}

\begin{table}[htbp]
\caption{Source spectrum fit with \texttt{vnei} model. The best fitted parameter values with the $1 \sigma$ uncertainties are listed for the two sector and one sector background models. The silicon and iron abundance are set to zero, and the sulfur abundance is set to 0.2.}
\begin{center}
\begin{tabular}{ l c c }
\hline\hline
Parameters & two sectors & one sector \\
\hline 
${kT_{e}}\,\mathrm(keV)$ & $0.97^{+0.21}_{-0.14}$ & $0.91^{+0.19}_{-0.13}$\\
$n_\mathit{e}\,t_{1},(10^{10}\,\mathrm{cm}^{-3}\mathrm{s})$ & $7.24^{+5.38}_{-2.65}$ & $6.81^{+5.75}_{-2.54}$\\
$Norm,(10^{-5})$ & $1.82^{+0.52}_{-0.49}$ & $2.02^{+0.63}_{-0.55}$\\
Oxygen & $ 0.16^{+0.20}_{-0.09}$ & $0.14^{+0.15}_{-0.07}$\\
Neon & $ 0.34^{+0.13}_{-0.11}$ & $0.27^{+0.12}_{-0.05}$\\
Magnesium & $ 0.14^{+0.12}_{-0.10}$ & $0.13^{+0.11}_{-0.10}$\\
near background in source & $0.98^{+0.14}_{-0.14}$ & $0.81^{+0.12}_{-0.11}$ \\
near background scale & $1.00^{+0.02}_{-0.02}$ & $1.00^{+0.01}_{-0.01}$ \\
\textit{C-statistic}& 9669 & 10173\\
\textit{$\chi^{2}_{P}$} & 18486 & 18575\\
dof & 18665 & 18665\\
goodness& 0.46 & 0.49\\
\hline
\end{tabular}
\end{center}
\label{tbl:source.vnei.nosi.nofe}
\end{table}

\section{Image Analysis}\label{sec:imganalysis}

\begin{figure*}[htb!]
  \centering
   \includegraphics[width=0.9\textwidth]
   {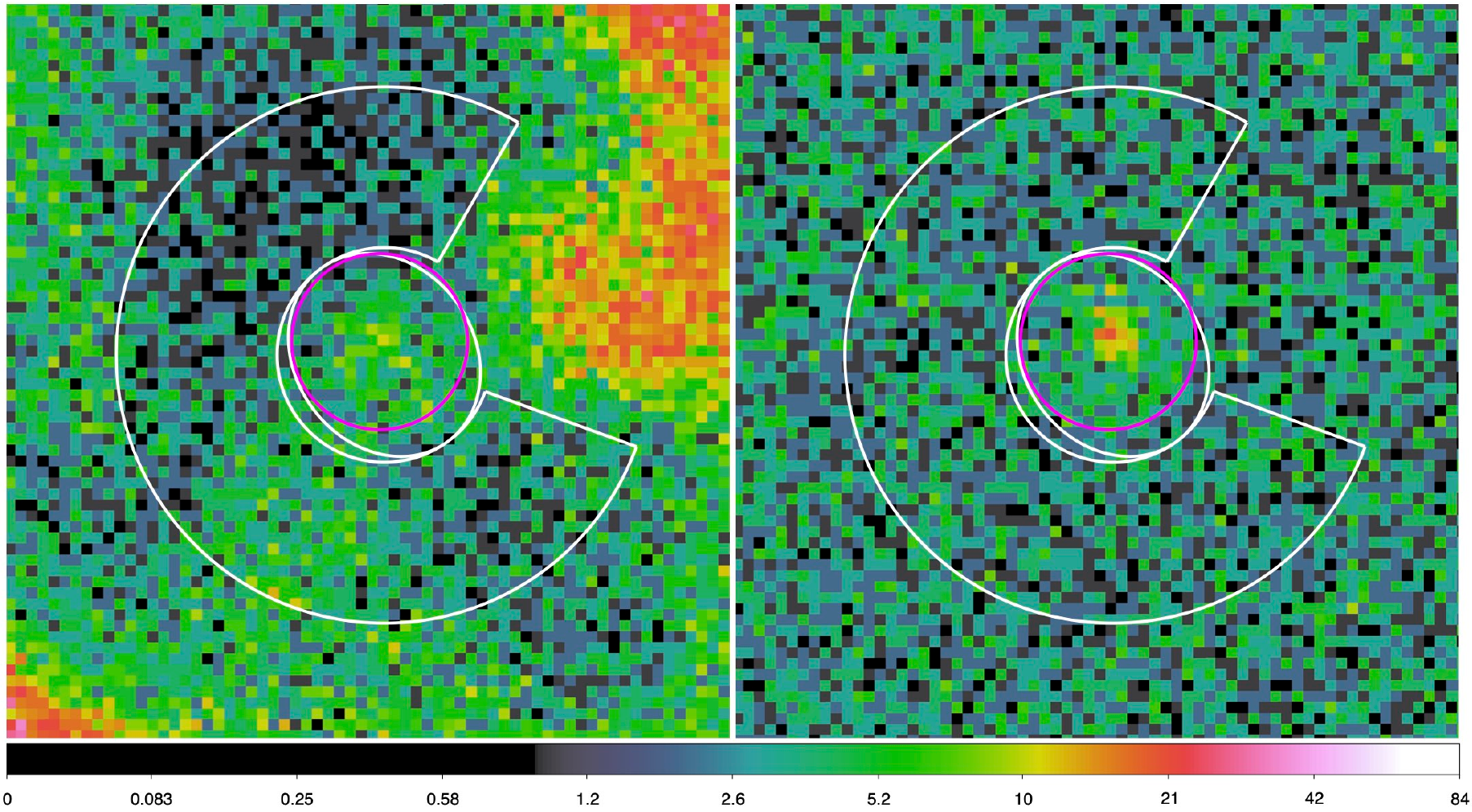}
   \caption{Left: Image of ten registered and merged observations. Right: Image created from a single instance of the 100 simulations. The source and near-background models in the simulation are taken from the last column in Table~\ref{tab:srcbb} for the one-sector near background. Outside the source ellipse, the near-background model is used. Inside the source ellipse, a blackbody point source is simulated ($kT$ and normalization free), with the near-background model rescaled using the ``near background in source'' correction factor from Table~\ref{tab:srcbb}. The image pixel size is $0.123\arcsec$.}
   \label{fig:countsdist.onesector.bb}
\end{figure*}

An image was constructed from those ObsIDs in Table~\ref{tbl:obslist} flagged as being used for the image analysis.  These ten ObsIDs were used in the analysis of \citet{long2019}, where the data were registered with positional uncertainties of $\sim 0.1\arcsec$.  We apply those positional corrections and use \texttt{dmmerge}\footnote{https://cxc.cfa.harvard.edu/ciao/ahelp/dmmerge.html} 
to merge the registered data sets.  The merged data are filtered 0.35-4.0\,keV and binned at 0.25 ACIS sky pixels ($0.123\arcsec$). The resulting image is presented in the left-hand panel of Figure~\ref{fig:countsdist.onesector.bb}. This figure also shows the outlines of
the source extraction region, the radial profile extraction region, and the one-sector near-background extraction region. 

\begin{figure*}[!tbh]
\begin{center}
\begin{tabular}{cc}
\includegraphics[width=0.45\textwidth]
   {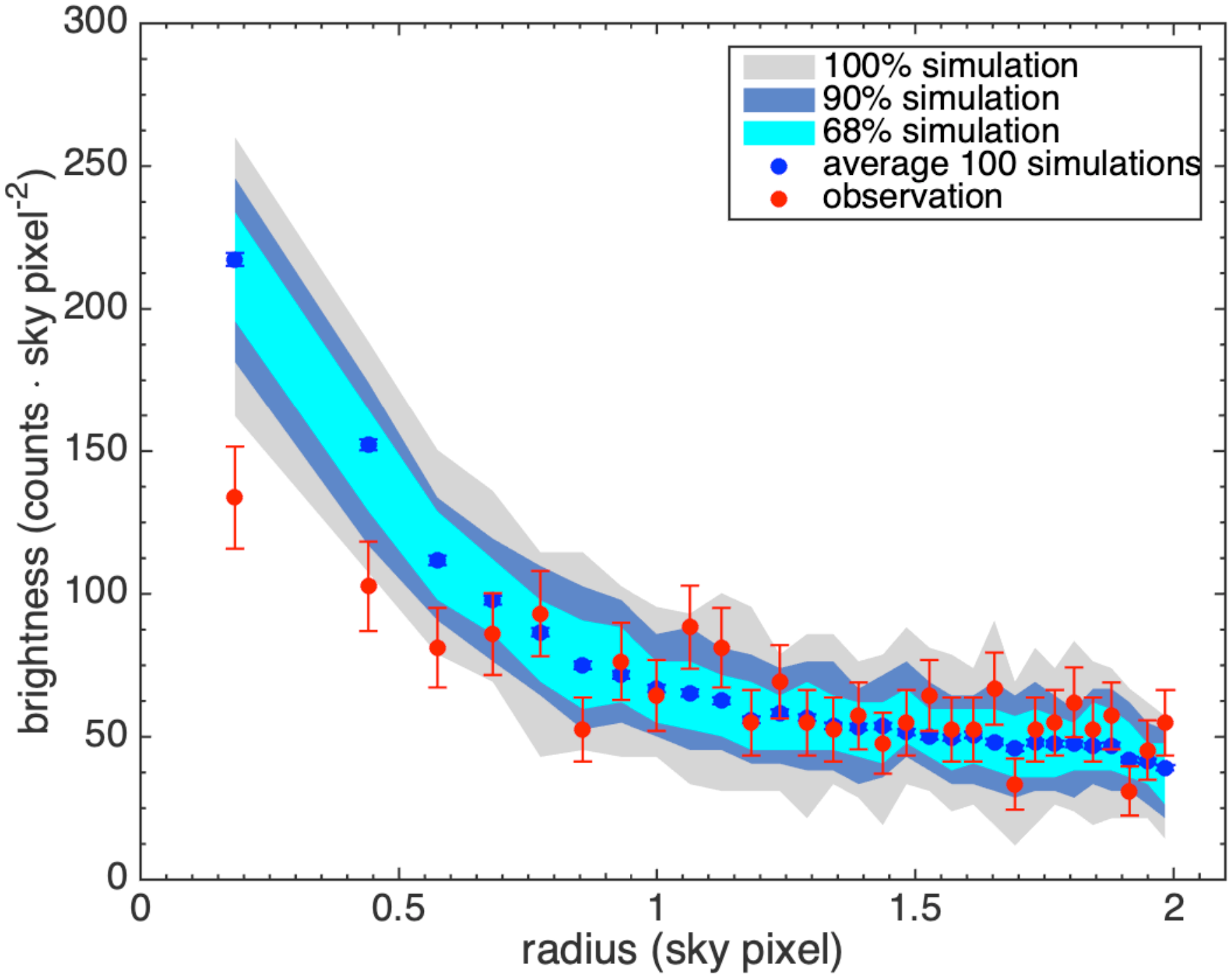}
\includegraphics[width=0.45\textwidth]
   {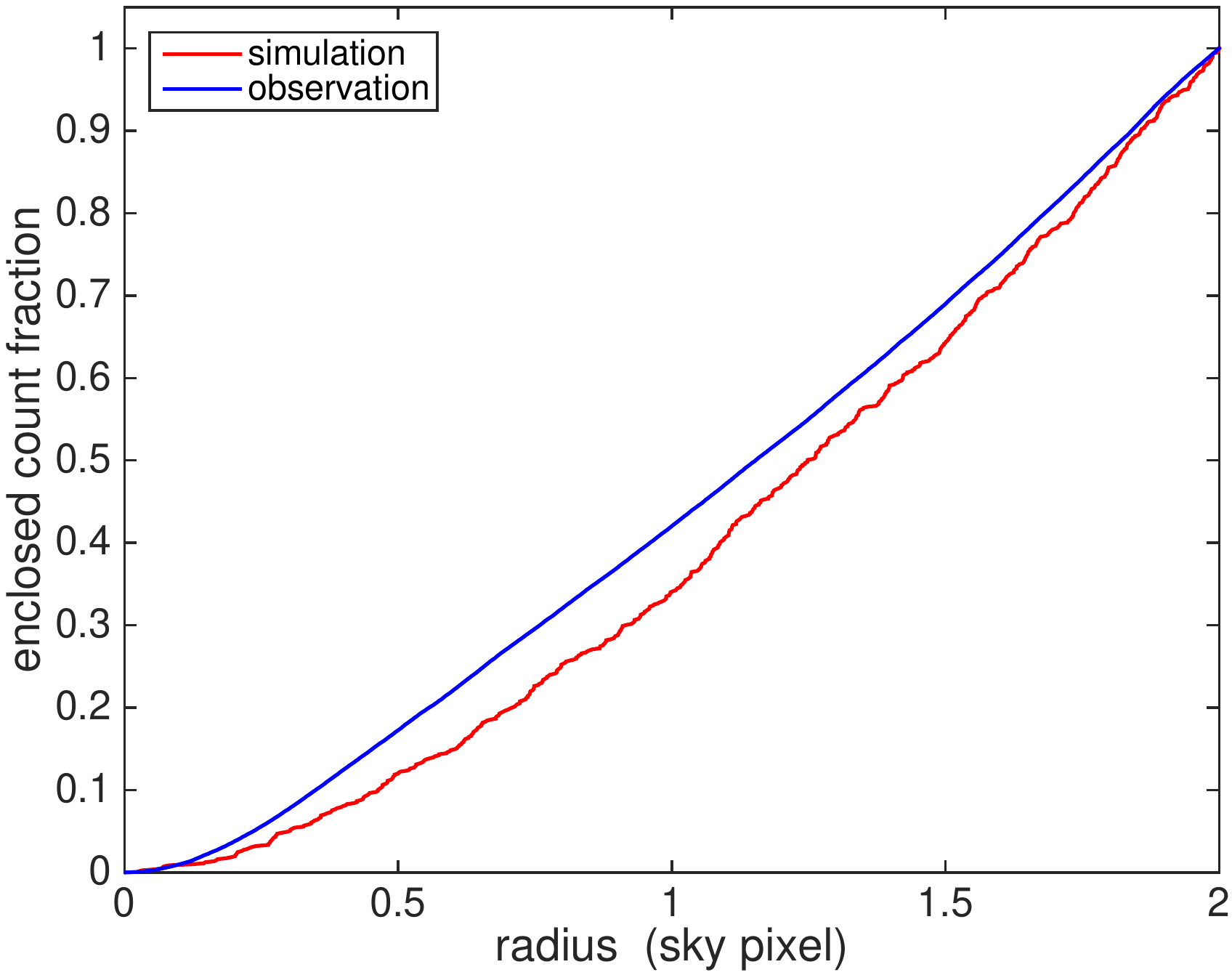}
\end{tabular}
  \caption{Results for one-sector near-background and blackbody model with $kT$ and normalization free. The simulated data are in blue, and the observed data are in red.  The simulated blackbody point source contributes 281 counts in the source region.
Left: Equal-area-binned radial surface brightness profiles.
Right: Naturally-binned cumulative surface brightness profiles normalized to one at 2.0 ACIS sky pixels (0.984\arcsec).
}
\label{fig:one.sector.profile.and.cumulative.profile.bb}
\end{center}
\end{figure*}

The X-ray image of the putative CCO does not appear to be consistent with a point source.  In order to test this quantitatively, we perform an imaging analysis to assess whether the observed distribution of counts is consistent with a point source superposed on a background. We performed simulations with \saotrace\footnote{https://cxc.cfa.harvard.edu/cal/Hrma/SAOTrace.html} to model a point source at the location of the putative CCO, and used \marx\footnote{https://space.mit.edu/cxc/marx/} to apply the detector response and to model the near-background. The position of the putative CCO was estimated by using \texttt{dmstat}\footnote{https://cxc.harvard.edu/ciao/ahelp/dmstat.html} to compute the centroid of data near the center of the source extraction region as shown in the left panel of Figure~\ref{fig:countsdist.onesector.bb}. This centroid, the position of the compact feature reported in \cite{vogt2018}, and the center of the source spectrum extraction region are listed in Table~\ref{tbl:img.centers}. This centroid is located $0.35\arcsec$ northeast of the reported CCO position. We extracted radial profiles from the centroid of the image within a radius of 2 sky pixels ($0.984 \arcsec$). The radial profile extraction region is the magenta circle shown in Figure~\ref{fig:countsdist.onesector.bb}.
The Kolmogorov-Smirnov two-sample test is applied to assess whether the radial distribution from the observed data is consistent with the simulated point-source plus background data.

Because the observations were carried out nearly on-axis, but with different pointings and roll angles, the PSF varies somewhat from observation to observation. In addition, the detector response varies with time because of the contamination buildup on the optical blocking filter, so the as-detected PSF and flux also vary with detector response. We therefore perform simulations for each of the ten ObsIDs listed as image analysis observations in Table~\ref{tbl:obslist}, applying the appropriate exposure time, pointing, roll, and detector response.

We generated 100 simulations of the stacked observed data.  Each ObsID was simulated 100 times with different seeds.  For each of the 100 sets of simulations, the simulated event lists for the ObsIDs were merged using \texttt{dmmerge}. Images were generated from the simulated stacked data, using \texttt{dmcopy} to filter for energies 0.35-4.0\,keV and binning to 0.25 ACIS sky pixels (0.123\arcsec). Each of the images has comparable count statistics as the stacked observations.  For the Kolmogorov-Smirnov tests, the 100 merged simulation event lists were merged and filtered to 0.35-4.0\,keV to provide a well sampled simulation model.

The local background within the source extraction region and near-background extraction regions is primarily due to diffuse E0102 emission which varies spatially.  The actual spatial distribution of E0102 emission within the source extraction region is not known. In the simulations, we approximate the background from E0102 emission as a spatially uniform distribution.  The near background  was estimated using \saotrace + \marx to simulate a spatially uniform background within a disk with radius 12\arcsec (large enough to encompass the source and near-background extraction regions). In the spectral analysis sections, we investigated two options for obtaining ``near background'': two-sector (Figure~\ref{fig:srcbkg}) and one-sector (Figure~\ref{fig:srcbkg.alternative}). The near background spectral models were used to estimate the local background in the source region, but the normalization of the near-background model fit to the source spectrum was allowed to vary independently from the normalization of the near-background model (simultaneously) fit to the near-background spectrum.
For the local background in the source region we take the near-background spectral model from Table~\ref{tbl:2.component.nearback} in \S\ref{sec:spectral.analysis}, modified by the ``near background in source'' factor in Table~\ref{tab:srcbb}.  
The detector plus blank sky background was simulated using a blank sky event list from CALDB\footnote{acis7sD2005-09-01bkgrndN0005.fits}; however, the blank sky and the detector backgrounds were minor contributions compared to the diffuse E0102 emission. 

\begin{table}[htbp]
\caption{Location of the compact feature and the centers of the ellipses for the spectral and imaging analyses}
\begin{center}
\label{tab:geocenter}
\begin{tabular}{l l l}
\hline\hline
centers & RA & DEC\\
\hline
compact feature        & 01h\,04m\,2.7s  & $-72^{\circ}\, 02\arcmin\, 00.2\arcsec$  \\
source ellipse region$^{\ddagger}$   & 01h\,04m\,2.75s & $-72^{\circ}\, 02\arcmin\, 00.14\arcsec$ \\
centroid of image       & 01h\,04m\,2.76s & $-72^{\circ}\, 01\arcmin\, 59.99\arcsec$ \\
\hline
\end{tabular}
\begin{tablenotes}
        \footnotesize
        \tablenotetext{\dagger} {Compact feature location reported in \cite{vogt2018}}
        \tablenotetext{\ddagger} {Center of the source spectral analysis region in \cite{vogt2018} and of source spectral analysis in this paper.}
\end{tablenotes}
\end{center}
\label{tbl:img.centers}
\end{table}

The point source was modeled in \saotrace using the \texttt{vnei} model (see \S\ref{subsubsec.vnei}, Table~\ref{tbl:source.vnei}) as the spectral model.  We also performed simulations in which the source model was based on a blackbody with the $kT$ and normalization set to the values in the right hand columns in Table~\ref{tab:srcbb}, which represent the best fit achieved with the blackbody model.
Although the models in Table~\ref{tab:srcbb} do not provide acceptable fits for a blackbody model, they allow for a more stringent limit on the possibility of a point source.

The simulated counts in the near-background extraction regions were matched to the observed counts in the near-background extraction regions in the merged observations (Table~\ref{tbl:counts.dist.vnei.bb}) by applying a scale factor to the simulation model. In fitting the spectrum for the source region, the model includes a source model, a near-background model, and a blank-sky sky plus detector background (see \S\ref{sec:spectral.analysis}). In simulating the source region, the normalization of the near-background model is adjusted according to the ``near background in source'' factor from the last column of Table~\ref{tab:srcbb}.  This factor (based on the spectral fitting) allows for the local background in the source region to have the same shape as in the near-background region, but with a different normalization.

The near-background simulation and the local background in the source region both used the near-background model, but with different normalizations. This necessitated running the simulations twice. First, for the simulations of the near-background region, a near-background model scale factor of 1.0 was used. Second, in simulating the background in the source region, the near-background model was scaled by the ``near background in source'' factor from the last column in Table~\ref{tab:srcbb}. In plots of images of the simulation, the data outside the source extraction region is from the first set of simulations, and the data for the source extraction region is from the second set of simulations.
Table~\ref{tbl:counts.dist.vnei.bb} compares the counts in the simulations to the counts in the observation.  The breakdown of the number of counts in the source region due to the point source, the near-background model, and the blank sky/detector background is given in Table~\ref{tbl:counts.decomp.vnei.bb}. 

We present here the blackbody point source case with the one-sector near background data; this examines the case of the weakest point source and highest background level based on the spectral fitting (see Table~\ref{tbl:counts.decomp.vnei.bb}).  The left panel of Figure~\ref{fig:countsdist.onesector.bb} shows the merged data and the source extraction and one-sector near-background region plotted. The right panel shows the results of a single instance of the corresponding simulations, to provide comparable count statistics. 
The other cases, blackbody source model plus two-sector near background and vnei source model with one- and two-sector backgrounds, are similar.  The spatial distribution of a point source is much more peaked and compact than the more diffuse emission observed in the source region. This is even more the case with a blackbody point source and two-sector near background, and \texttt{vnei} source with one-sector or two-sector near background.

\begin{table*}[htb!]
\caption{Source and Background Counts}
\begin{center}
\begin{tabular}{| l | c | c| c |}
\hline
 &  & vnei & blackbody\\
Region & observation & simulation & simulation \\
 & [counts] & [counts] & [counts] \\
\hline
two-sector near-background & 1784  & 1784  & 1784 \\
source ellipse & \phantom{0}983 & \phantom{0}981 & \phantom{0}979\\
\hline
one-sector near-background & 3248  & 3248 & 3248 \\
source ellipse & \phantom{0}983 & \phantom{0}987 & \phantom{0}984\\
\hline
\end{tabular}
\end{center}
\label{tbl:counts.dist.vnei.bb}
\end{table*}

\begin{table}[htb!]
\caption{Source Region Decomposition: Simulation \textit{vs.} Observed}
\begin{center}
\begin{tabular}{| l | c c |}
\hline
two-sector background
 & \texttt{vnei} & blackbody \\
 & [counts] &  [counts] \\
\hline
scaled background & \phantom{0}552 & \phantom{0}669 \\
source & \phantom{0}425 & \phantom{0}306 \\
S3/sky background & \phantom{000}4 & \phantom{000}4 \\
\hline
\hline
one-sector background
 & \texttt{vnei} & blackbody \\
 & [counts] &  [counts] \\
\hline
scaled background & \phantom{0}592 & \phantom{0}699 \\
source & \phantom{0}391 & \phantom{0}281 \\
S3/sky background & \phantom{000}4 & \phantom{000}4 \\
\hline
\end{tabular}
\end{center}
\label{tbl:counts.decomp.vnei.bb}
\end{table}

Radial profiles were constructed from the simulated event lists and the merged observation event list by sorting events by distance from the point source position.  Figure~\ref{fig:one.sector.profile.and.cumulative.profile.bb} displays the radial profile for the blackbody point source and one-sector near background model simulations. The left panel shows the radial profiles with equal-area binning to provide similar count statistics within each radial bin. The right panel shows the cumulative radial profiles based on the merging of the 100 simulations. The cumulative radial profile uses a ``natural binning'' for the radial data. Histograms are constructed in which the value increases by one at the radius of each event encountered; this is done for both the observational data and the merged simulations.  Both histograms are normalized to have a maximum value of one for use in the Kolmogorov-Smirnov two-sample tests discussed below.
The radial profile in the simulation is notably more peaked than the profile from the observation. This is also the case with a blackbody point source and two-sector near background, and even more so with a \texttt{vnei} source with one- or two-sector near backgrounds. In each case, the profiles differ systematically by more than 1$\sigma$ within a radius of $\lesssim 0.7$\,ACIS sky pixels.  
Even in the most optimistic case of the blackbody point source model with the one-sector background model (which has the highest relative background level), the radial profile is significantly more peaked in the simulation than in the observation.

To apply the Kolmogorov-Smirnov two-sample test, the value $D$ (maximum absolute difference between the cumulative distributions) was calculated for each case; this corresponds to the maximum vertical distance between the blue and red histograms in the right hand panel of Figure~\ref{fig:one.sector.profile.and.cumulative.profile.bb}
for the case of a blackbody point source and a background based on the one-sector near background.
The Kolmogorov-Smirnov test critical value $D_{0.999}(m,n)$ (for a 0.999 significance level) is $D_{0.999}(m,n) = 1.95\sqrt{(m+n)/(mn)}$ where $m$ and $n$ are the sample sizes for the two histograms \citep{wall2012,smirnov1948}.  If $D$ exceeds $D_{0.999}(m,n)$, the distributions are inconsistent at the 99.9\% level. The results for the vnei and blackbody source models, and for the one-sector and two-sector near-sky background models, are given in Table~\ref{tbl:kstest}. The Kolmogorov-Smirnov test shows that the observed distribution is inconsistent with the point-source plus background model at the $>99.9$\% level for each combination of point-source spectrum (blackbody and \texttt{vnei}) and near-background model (one-sector and two-sector near background). 

To compare the observed and simulated radial profiles, we calculated a $\chi^2$ value using the 30 bins in the observed radial profile as the data, the average value of the 100 simulated radial profiles as the model, and weighted by the error of the data. We evaluated the $\chi^2$ distribution for 30 degrees of freedom. The P-value was obtained by integrating this $\chi^2$ distribution over values larger than the $\chi^2$ value obtained from the radial profiles. The results for the vnei and blackbody source models, and for the one-sector and two-sector near-sky background models, are given in Table~\ref{tbl:chi2}. The P-values show that the observed radial profiles are not consistent with the simulated radial profiles at confidence level exceeding 99.9\%.  This is consistent with the results from the Kolmogorov-Smirnov test.

\begin{table}[htb]
\caption{The sample size and Kolmogorov-Smirnov test statistics.}
\begin{center}
\begin{tabular}{| l | c c c c |}
\hline
Near-Background &\multicolumn{4}{c|}{Point Source: vnei}\\
\hline
 & m & n & $D_{0.999}(m,n)$ & $D$\\
\hline
two-sector& 86499 & 809  & 0.069 & 0.16\\
one-sector& 86159 & 809  & 0.069 & 0.15\\
\hline
\hline
Near-Background &\multicolumn{4}{c|}{Point Source: blackbody}\\
\hline
& m & n & $D_{0.999}(m,n)$ & $D$\\
\hline
two-sector& 83766 & 809 & 0.069 & 0.10 \\
one-sector& 83831 & 809 & 0.069 & 0.08 \\
\hline
\end{tabular}
\end{center}
\label{tbl:kstest}
\end{table}

\begin{table}[htb]
\caption{The $\chi^{2}$ and P-value for radial profiles.}
\begin{center}
\begin{tabular}{| l | c c c|}
\hline
Near-Background &\multicolumn{3}{c|}{Point Source: vnei}\\
\hline
 & $\chi^{2}$ & P-value & Dof\\
\hline
two-sector& 178.91 & $4.02\times10^{-23}$  & 30\\
one-sector& 144.46 & $6.39\times10^{-17}$  & 30\\
\hline
\hline
Near-Background &\multicolumn{3}{c|}{Point Source: blackbody}\\
\hline
& $\chi^{2}$ & P-value & Dof\\
\hline
two-sector& 74.68 & $1.12\times10^{-5}$ & 30\\
one-sector& 61.93 & $5.33\times10^{-4}$ & 30\\
\hline
\end{tabular}
\end{center}
\label{tbl:chi2}
\end{table}

\begin{figure*}[!tbh]
\begin{center}
\begin{tabular}{cc}
\includegraphics[width=0.45\textwidth]
   {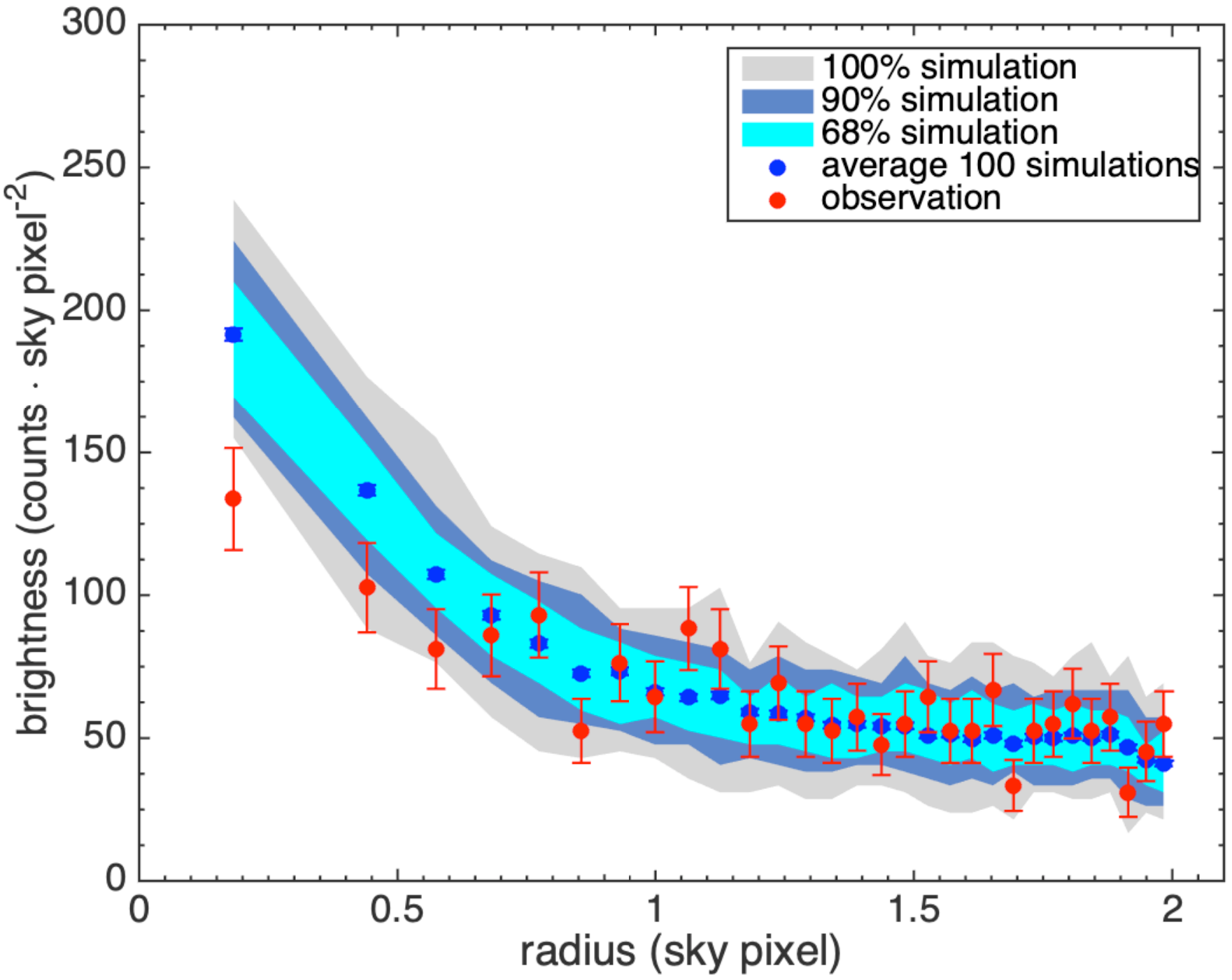} &
   \includegraphics[width=0.45\textwidth]
   {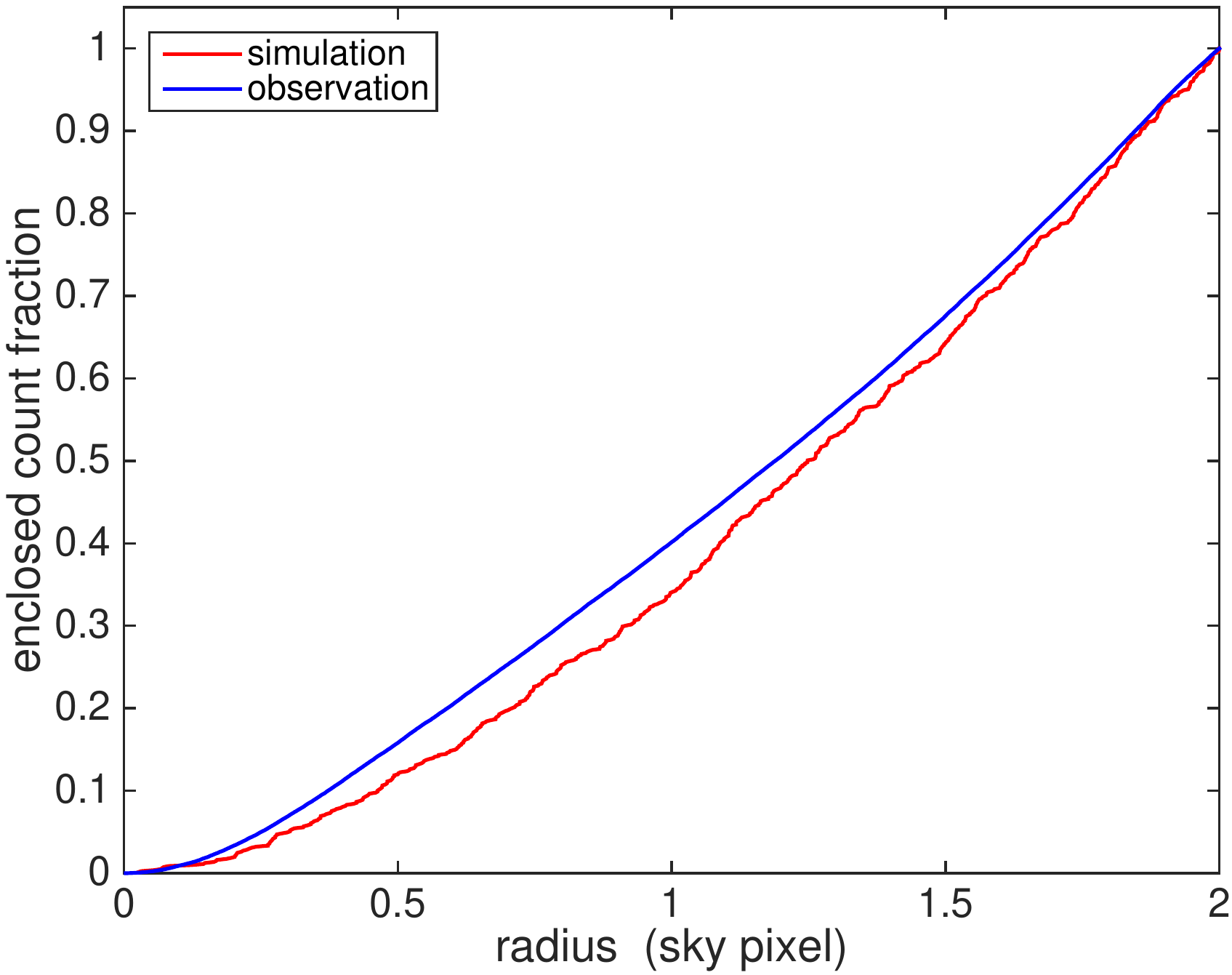}
\end{tabular}
  \caption{Surface brightness radial profiles and cumulative radial profiles (observation in red, simulated data in blue), for the brightest point source that could not be rejected at the 99.73\% confidence level. Blackbody point source and one-sector near background. The simulated blackbody point source contributes 241 counts in the source region.
     Left: Equal-area-binned radial surface brightness profiles.
     Right: Naturally-binned cumulative surface brightness profiles normalized to one at 2.0 ACIS sky pixels (0.984\arcsec).
     }
\label{fig:one.sector.profile.and.cumulative.profile.bb.upper.limit}
\end{center}
\end{figure*}

Given the above results, a natural question is: what is an upper limit for a blackbody point source such that it would not be excluded by the Kolmogorov-Smirnov test at the 3$\sigma$ level. To examine this, we considered the one-sector near background spectral model and a blackbody point source model with parameters taken from the ``normalization and temperature free'' case in Table~\ref{tab:srcbb}.  We redid the simulations and varied the scaling of the near-background in the source region and the normalization of the blackbody (but not varying the $kT$) under the constraint that the total counts still match the observed counts in the source extraction region.
Figure~\ref{fig:one.sector.profile.and.cumulative.profile.bb.upper.limit} shows the radial profiles (left panel) and the cumulative radial profiles (right panel) for the modified simulations and the observed data. For a $3\sigma$ (99.73\%) criterion, the critical Komogorov-Smirnov criterion becomes 
$D_{0.9973}(m,n)=1.82\sqrt{(m+n)/(m n)}$ \citep{smirnov1948}.  For the simulation, $m=82980$ while the observation has $n=809$, resulting in a critical $D_{0.9973} = 0.0643$. For a simulation with the blackbody source providing 241 counts in the source region, the $D = 0.0624$ providing an estimated $3\sigma$ upper limit of 241 blackbody point source counts in the source region. 

\section{Discussion}\label{sec:discussion}

An X-ray point source with the blackbody spectral properties specified in \cite{vogt2018} is ruled out by our spectral analysis and the analysis of \cite{hebbar2019}.
\cite{hebbar2019} explored a variety of models including blackbody, power-law, unmagnetized and magnetized NS atmosphere models in different combinations, see their paper for the details.
They were able to achieve an acceptable fit with a blackbody~+~power-law model or a NS C atmosphere model. The addition of a power-law component was necessary to model the emission at higher energies that is not reproduced by a single blackbody model, similar to what we show in Figure~\ref{fig:srcbbnewnb}.    A blackbody~+~power-law model provides an acceptable fit with a temperature of $kT=0.18$~keV, the power-law index frozen to 2.0,  $N_{H},_{SMC}=5.6 \times 10^{21} \mathrm{cm}^{-2}$, and luminosities in the blackbody and power-law components of 
$L_{BB}=6.0\times10^{33}~{\mathrm{erg s^{-1}}}$ and $L_{pl}=1.4\times10^{33}~{\mathrm{erg s^{-1}}}$.
The NS H atmosphere models were rejected for not reproducing the emission at high energies.
A NS C atmosphere model with an effective temperature of $kT=0.26$~keV, a magnetic field of $B=10^{12}$~G , $N_{H},_{SMC}=9.0 \times 10^{21} \mathrm{cm}^{-2}$, and a luminosity of  $L\sim1.0\times10^{34}~{\mathrm{erg s^{-1}}}$ also provides an acceptable fit.
\cite{hebbar2019} note that the inferred temperature and luminosity for this source are higher than most NSs, in particular the thermal luminosity is higher than most rotation-powered pulsars and indicates that the origin of the emission might be related to the high magnetic field of the source.  They suggest that this source might be an object with properties that position it  between the known CCOs with relatively low magnetic fields and luminosities  and the magnetars with relatively high magnetic fields and luminosities.  One possibility is that the magnetic field was buried by fallback accretion after the SN explosion \citep{ho2011} and is beginning to emerge such that the inferred field of $B=10^{12}$~G is larger than that of the CCOs but less than that of the magnetars.  One difficulty with both of these models is that they require a large value of the column density within the SMC along the line of sight to E0102 of  $N_{H},_{SMC}=5.7\times 10^{21} \mathrm{cm}^{-2}$ or $N_{H},_{SMC}=9.0\times 10^{21} \mathrm{cm}^{-2}$ which are 5-10 times larger than the values derived from spectral fits to the emission from the SNR in \cite{plucinsky2017} and \cite{alan2019}.  For these models to be correct, there must be a significant density enhancement along the line of sight to this compact feature that is not present in the lines of sight to the adjacent regions.

We have shown that a thermal NEI spectral model, similar to the surrounding background regions and the ejecta emission in this SNR, fits the data well. The abundances, and hence the line emission, are significantly lower in our source models than in our near background models.  However, the abundances
are not well constrained given that the near background dominates the source spectrum in this region and the limited statistical quality of the spectra.  The relatively lower abundances in the source region compared to the near 
background region could indicate a real variation in the abundances from the near background region to the source region or a difference in the amount of ejecta that have been shocked to X-ray emitting temperatures, as hinted at
by our second set of vnei fits with a relatively low ionization timescale.  We can not exclude the possibility that there is nonthermal emission from either the forward or reverse shock that is contributing in this region
and reducing the contrast between the line emission and the continuum.  We also can not exclude the possibility that there is nonthermal emission associated with a nebula around a point source fainter than the flux described in \citet{vogt2018}. But we have shown that
thermal models fit the spectral data as well as the models
explored by \cite{hebbar2019}.

More importantly, we have shown that the spatial distribution of the counts is not consistent with that of a point source. Therefore, the simplest explanation for this compact feature is another clump of SN ejecta associated with the reverse shock in E0102.
If there is a point source in this region, it must be significantly fainter than the values quoted in \cite{vogt2018} and \cite{hebbar2019}.  The faintest point source which we considered in our spectral analysis with the blackbody model and the one sector background model had a luminosity of $L(0.35-4.0~{\mathrm{keV}})\sim3.90\times10^{33}~{\mathrm{erg\, s^{-1}}}$, a factor of 2.7 times less luminous than the \cite{vogt2018} model. 
A point source with this luminosity is inconsistent with the image data at the 99.9\% confidence level and was also rejected in the spectral analysis. There may still be a point source embedded in this region of diffuse emission but its flux must be significantly lower than the values reported in \cite{vogt2018} and  \cite{hebbar2019}.
The brightest point source that is still consistent with the radial distribution of counts described in  \S\ref{sec:imganalysis} would have a luminosity of  $L(0.35-4.0~{\mathrm{keV}})\sim3.34\times10^{33}~{\mathrm{erg s^{-1}}}$. Any point source more luminous than this is rejected at the $3\sigma$ confidence level. 
Such a luminosity is still within the range of the known CCOs but given that this is our $3\sigma$ upper limit if there were a point source in this region its luminosity is most likely less than this value
. 

Another argument against the CCO explanation for this compact feature is its location and the inferred transverse velocity for the NS.
If we assume the explosion center for the SNR adopted in \cite{long2019}, the position of this compact feature would imply a transverse velocity of $v\sim900-1300~{\mathrm {km~s^{-1}}}$ for an assumed range of ages of 2700-1800~yr. The highest transverse velocity measured for a CCO is $v=763\pm73~{\mathrm {km~s^{-1}}}$ for RX~J0822-4300 in the Puppis A SNR \citep{mayer2020}. All of the other known CCOs are located closer to the presumed center of explosion of their respective SNRs and have significantly lower inferred transverse velocities. Pulsars have an average transverse velocity of $v\sim250~{\mathrm {km~s^{-1}}}$ 
\citep{hobbs2005,verbunt2017} but a handful of pulsars have velocities larger  than $1,000~{\mathrm {km~s^{-1}}}$. If the compact feature were indeed the NS from the SN explosion it would have the highest transverse velocity of any CCO and would have one of the highest transverse velocities for a NS.

Given that the radial distribution of the counts is not consistent with a point source for \chandra, the spectrum is fit equally well with a thermal NEI model, and the position of the compact feature implies a large transverse velocity from the center of the SNR, the most likely explanation for this feature is a clump of ejecta heated by the reverse shock of the SNR.
The prospects for detecting a point source with a luminosity less than $L(0.35-4.0~{\mathrm{keV}})\sim3.34\times10^{33}~{\mathrm{erg s^{-1}}}$ or reducing the upper limit on the flux of a point source in this region with ACIS are not promising as the low energy response continues to decline. However, a dedicated, deep observation of this region with the {\it High Resolution Camera} (HRC) on \chandra\, should produce a radial profile that is more constraining for the existence of a point source given the superior spatial resolution and low energy sensitivity of the HRC compared to ACIS. We are planning such an observation for the coming year.

\section{Conclusions}\label{sec:conclusions}

We have conducted a spectral and image analysis of the archival \chandra data of the compact feature in the SMC SNR E0102 that has been suggested to be the first extragalactic CCO detected. We have used appropriate, time-dependent responses that account for the decrease in the ACIS effective area in our spectral analysis. We extracted background spectra from two different regions with significantly different background levels and generated background models for each region.  We have fit the unbinned spectra to preserve the spectral resolution of the instrument, preserving the sensitivity to line-like features.  We have fit the source and background spectra simultaneously with the two different background models and show that our results are not sensitive to the background model selected.  Our spectral analysis excludes a single blackbody model as more than 99\% of the simulated spectra have a fit statistic less than that of the data.  
We find that a thermal NEI model similar to the models used to fit the reverse shock-heated ejecta in E0102 fits the data well. We find two classes of NEI models fit the data equally well, one with $kT\sim0.79$~keV, an ionization timescale of 
$\sim3\times10^{11}\,\mathrm{cm}^{-3}\mathrm{s}$, and marginal evidence for enhanced abundances of O and Ne and the other with a temperature of $kT\sim0.91$~keV, an ionization timescale of  $\sim7\times10^{10}\,\mathrm{cm}^{-3}\mathrm{s}$, and abundances consistent with local ISM values. The limited statistics in the source spectrum do not allow us to distinguish between these classes of NEI models.
We extracted a radial profile of counts centered on the position of the compact feature and compared it to simulations of a point source embedded in a region with background levels determined by the two different background regions.  We find that the radial profile of the counts is not consistent with that expected for a point source with \chandra and may be rejected at the 99.9\% confidence level. The most likely explanation for this compact feature is a clump of O and Ne rich ejecta associated with the reverse shock. 

\acknowledgments
L.X. acknowledges support from \chandra Guest Observer grant GO9-20068X and Smithsonian Institution Scholarly Studies grant 40488100HH00201. T.J.G. and P.P.P. acknowledge support under NASA contract NAS8-03060 with the \chandra X-ray Center.
The authors thank the anonymous referee for suggestions which significantly improved this paper. The scientific results reported in this article are based on data obtained from the \chandra Data Archive, and observations made by the \chandra X-ray Observatory. This research has also made use of software provided by the \chandra X-ray Center (CXC) in the application package CIAO, and the NASA Astrophysics Data System (ADS).

\clearpage
\vfill\eject
\bibliographystyle{yahapj}
\bibliography{references}

\appendix
\counterwithin{figure}{section}

\section{Supplemental Spectral Analysis Material}
\label{appdx:suppl.spectral.analysis}
\begin{figure}[htb!]
   \centering
   \includegraphics[width=0.45\textwidth]
   {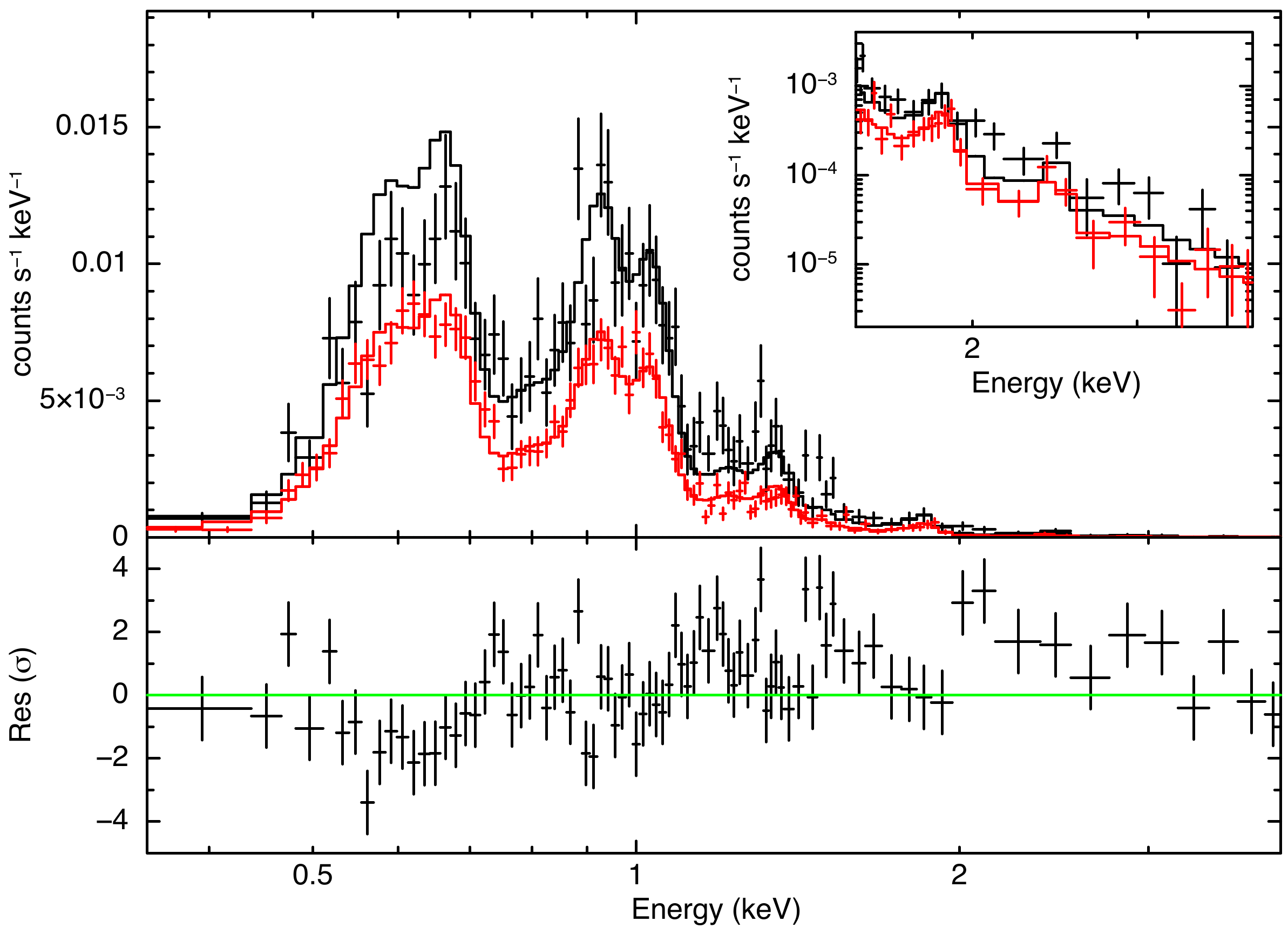}
   \caption{Source spectrum fit with the two sector near background model. The black points are the source data and the black line is the fitted near background model with only the normalization free. The red points and line are the near background data and model respectively.
   }
   \label{fig:scale.nearback}
\end{figure}

\begin{figure}[htb!]
  \centering
   \includegraphics[width=0.45\textwidth]
   {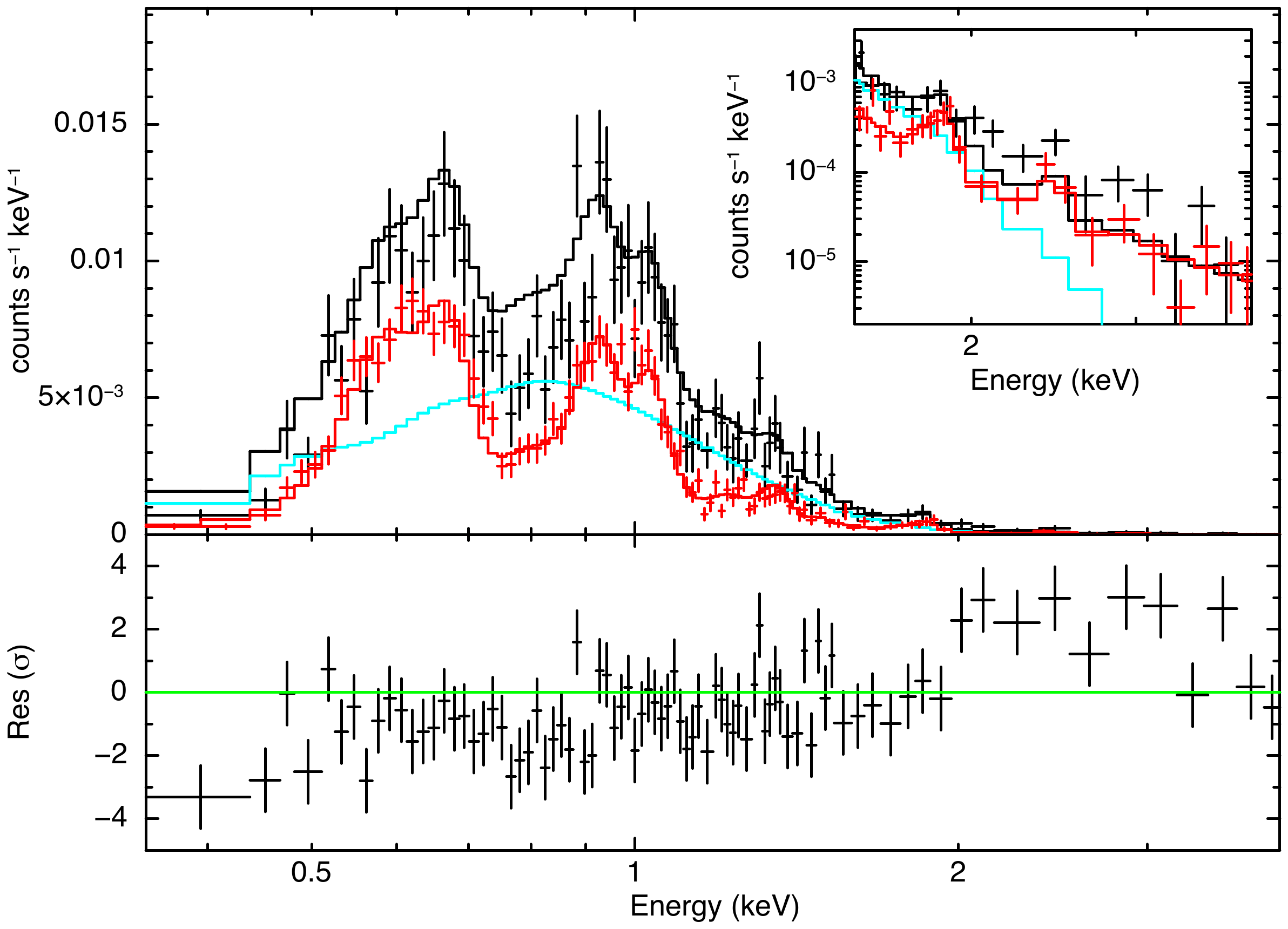}
   \caption{Source spectrum fit with a blackbody model and the two sector near background model. The blackbody parameters are fixed to the values in \cite{vogt2018} and the normalization of the near-background component (linked for the source and near background models) is allowed to vary.  The black points and line are the source data and model. The red points and line are the near-background data and model. The cyan line is the blackbody component in the source model. The blackbody model overpredicts the source spectrum in the 700-900~eV range and underpredicts above 2.0~keV.}
   \label{fig:srcbb}
\end{figure}

\begin{figure}[htb!]
  \centering
   \includegraphics[width=0.45\textwidth]
   {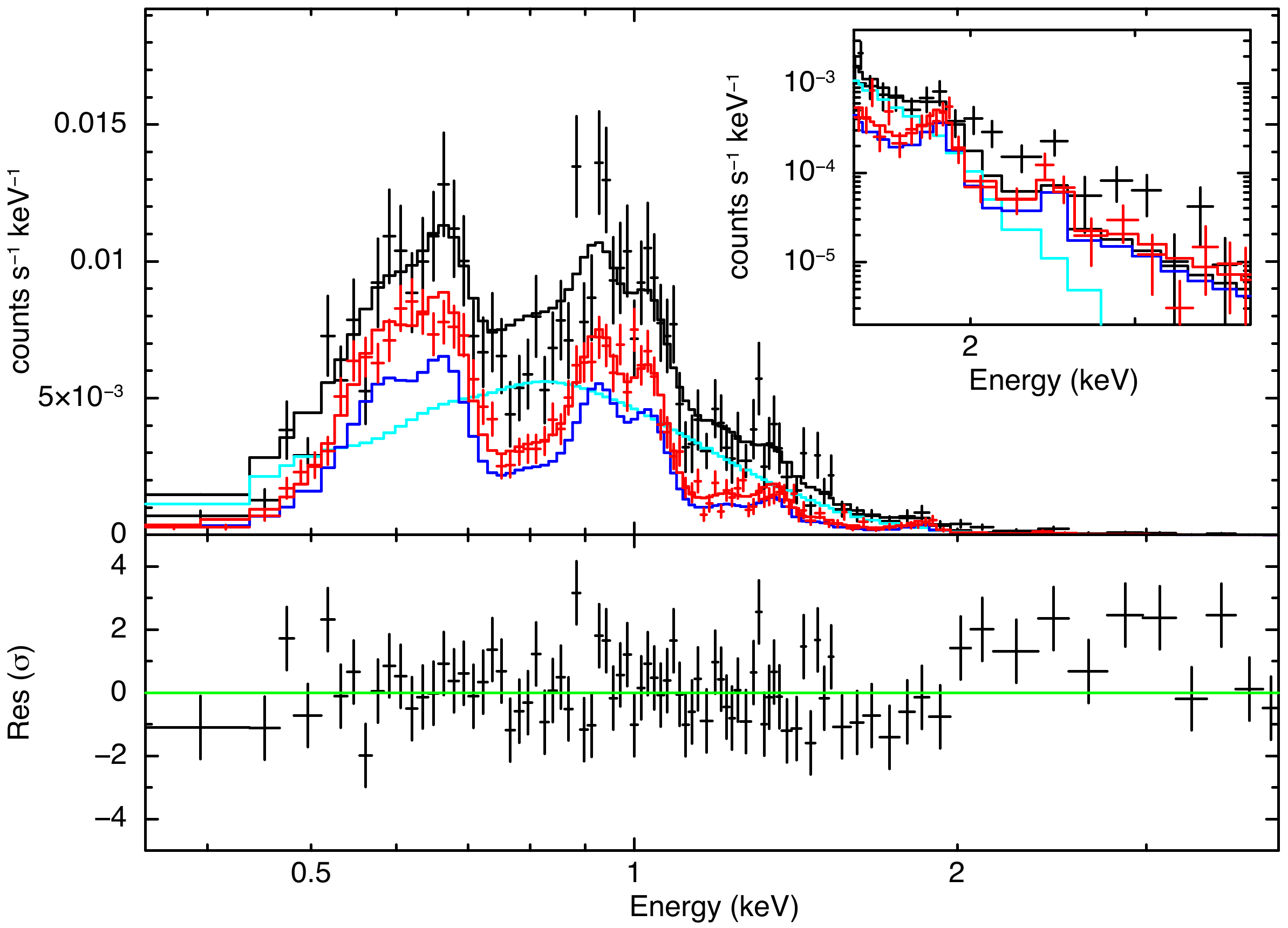}
   \caption{Source spectrum fit with a blackbody model and the two sector near background model. The blackbody parameters are fixed to the values in \cite{vogt2018} and the normalizations of the near background component in the  source and near background models are allowed to vary independently.
   The black points and line are the source data and model. The red points and line are the near background data and model. The cyan line is the blackbody component in the source model. The blue line is the near background component in the source model.}
   \label{fig:srcbbfnb}
\end{figure}

\begin{figure}[htb!]
  \centering
   \includegraphics[width=0.45\textwidth]
   {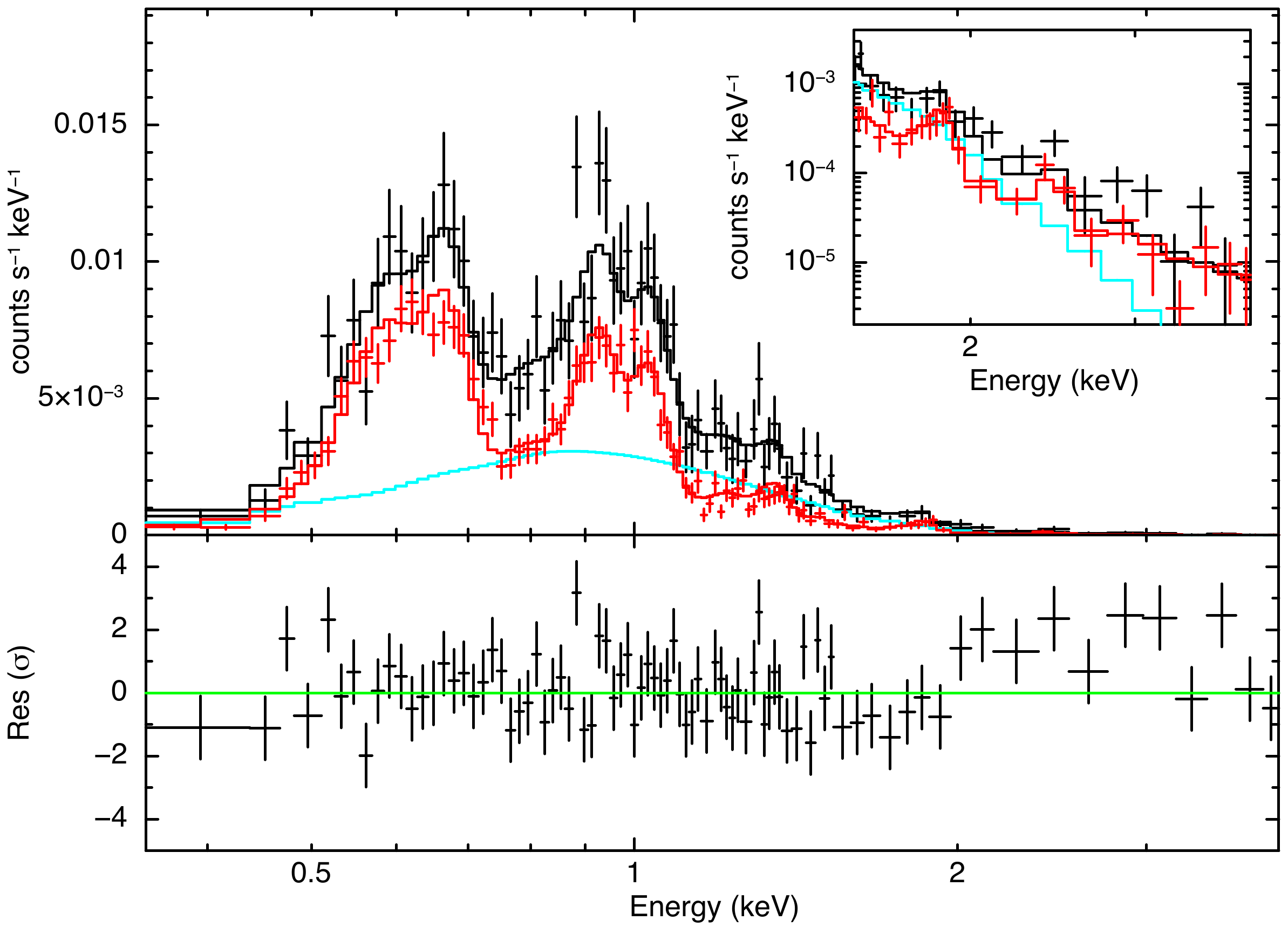}
   \caption{Source spectrum fit with a blackbody model and the two sector near background model. The temperature and normalization of the blackbody component  and the normalization of the near background
    component 
    (linked for the source and near background models) are free in the fit. The black points and line are the source data and model. The red points are the near background data and model. The cyan line is the blackbody component in the source model.}
   \label{fig:srcbbfp}
\end{figure}

\begin{figure}[htb!]
  \centering
   \includegraphics[width=0.45\textwidth]
   {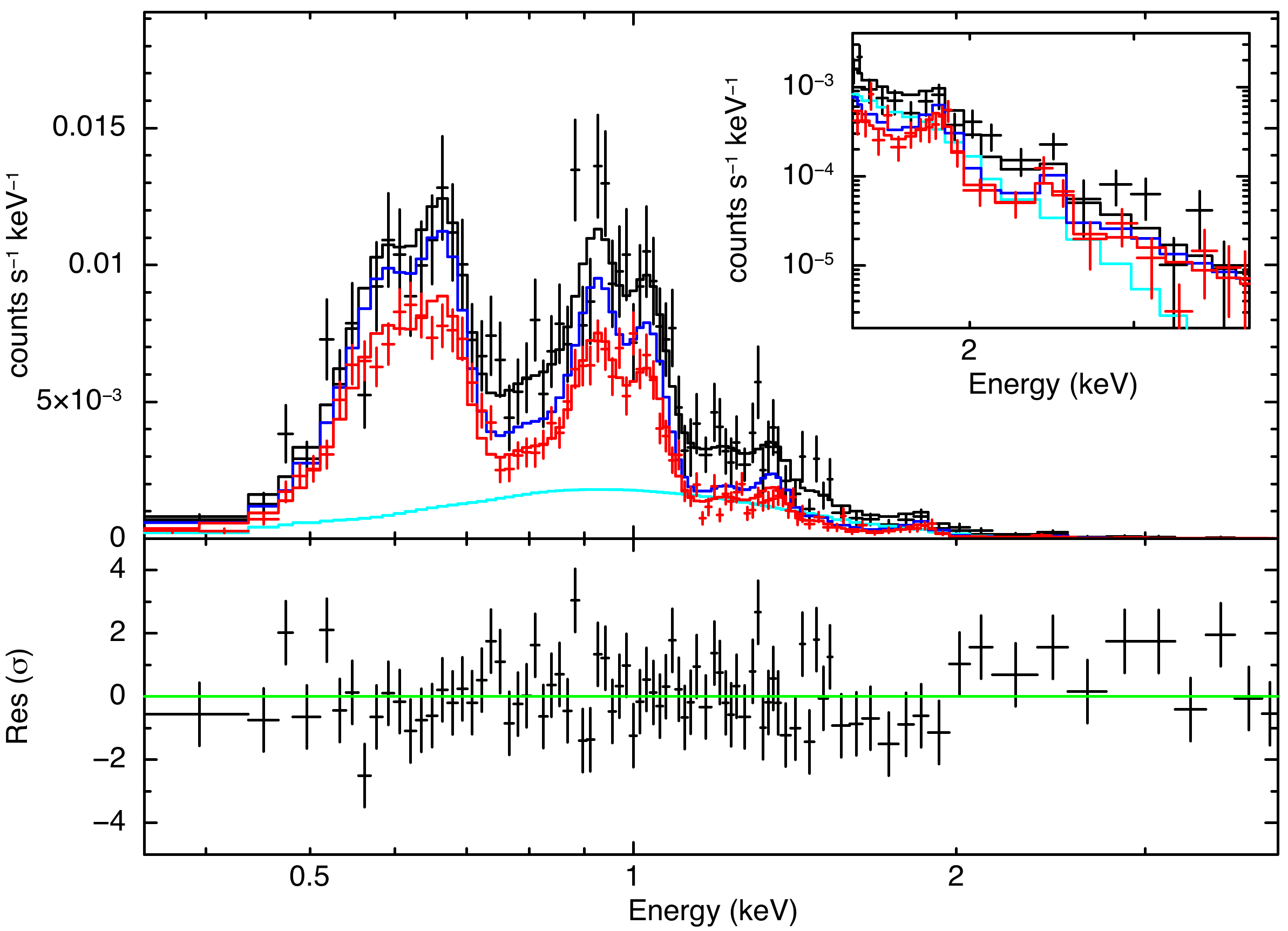}
   \caption{Source spectrum fit with a blackbody model and the two sector near background model. The temperature and normalization of the blackbody component  are free in the fit and the normalizations of the near background component in the  source and near background models are allowed to vary independently. The black points and line are the source data and model. The red points and line are the near background data and model. The cyan line is the blackbody component in the source model. The blue line is the near background component in the source model.}
   \label{fig:srcfbbfnb}
\end{figure}

\begin{figure}[htb!]
  \centering
   \includegraphics[width=0.45\textwidth]
   {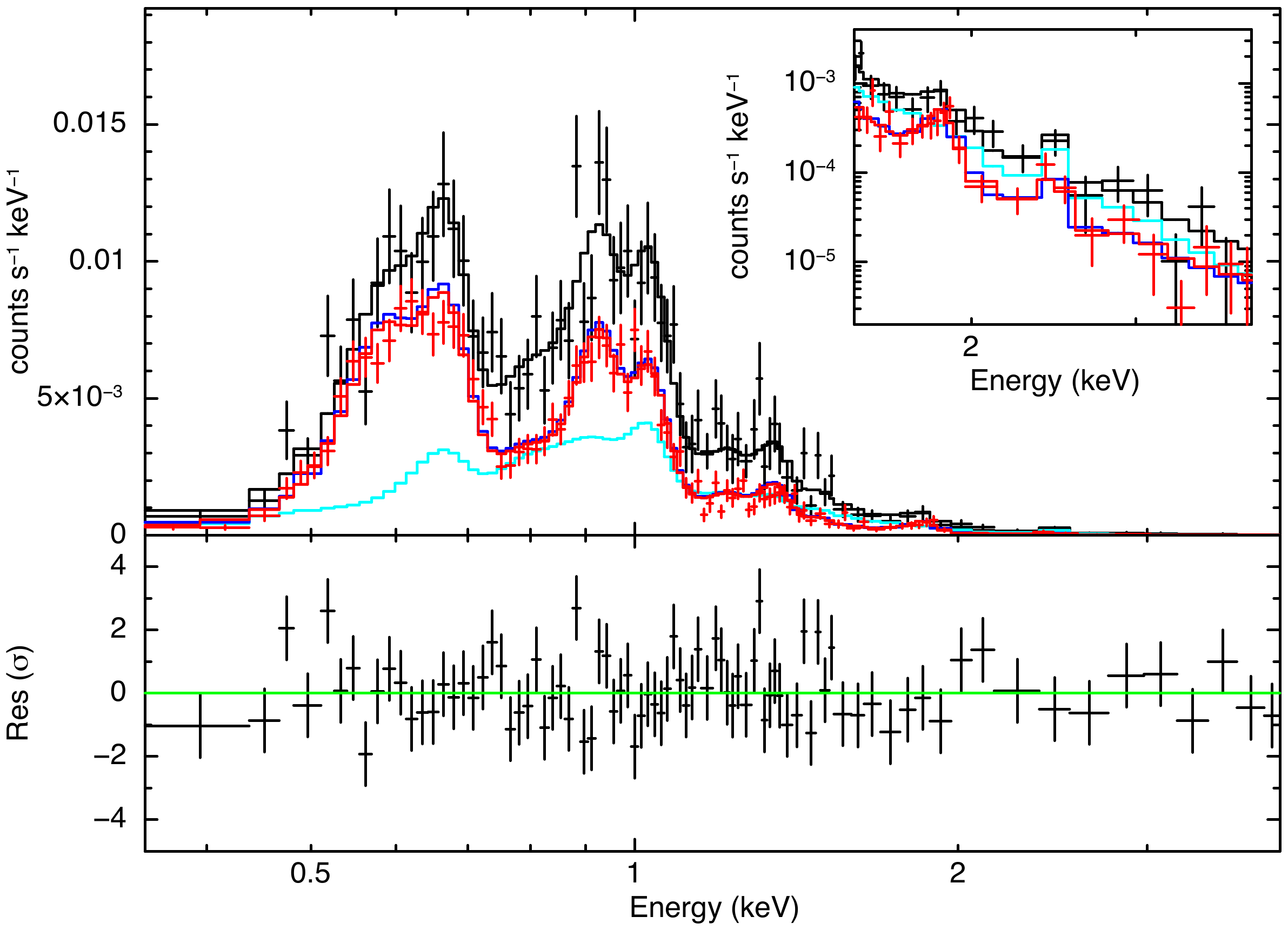}
   \caption{Source spectrum fit with a \texttt{vnei} model 
   ($n_\mathit{e}\,t \sim3.3\times10^{11}\,\mathrm{cm}^{-3}\mathrm{s}$)
   and the two sector near background model. The black points and line are the source data and model. The red points and line are the near background data and model. The cyan line is the \texttt{vnei} component in the source model. The \texttt{vnei} model parameters are listed in Table~\ref{tbl:source.vnei}.}
   \label{fig:srcvnei.free.fe.2}
\end{figure}

\begin{figure}[htb!]
  \centering
   \includegraphics[width=0.45\textwidth]
   {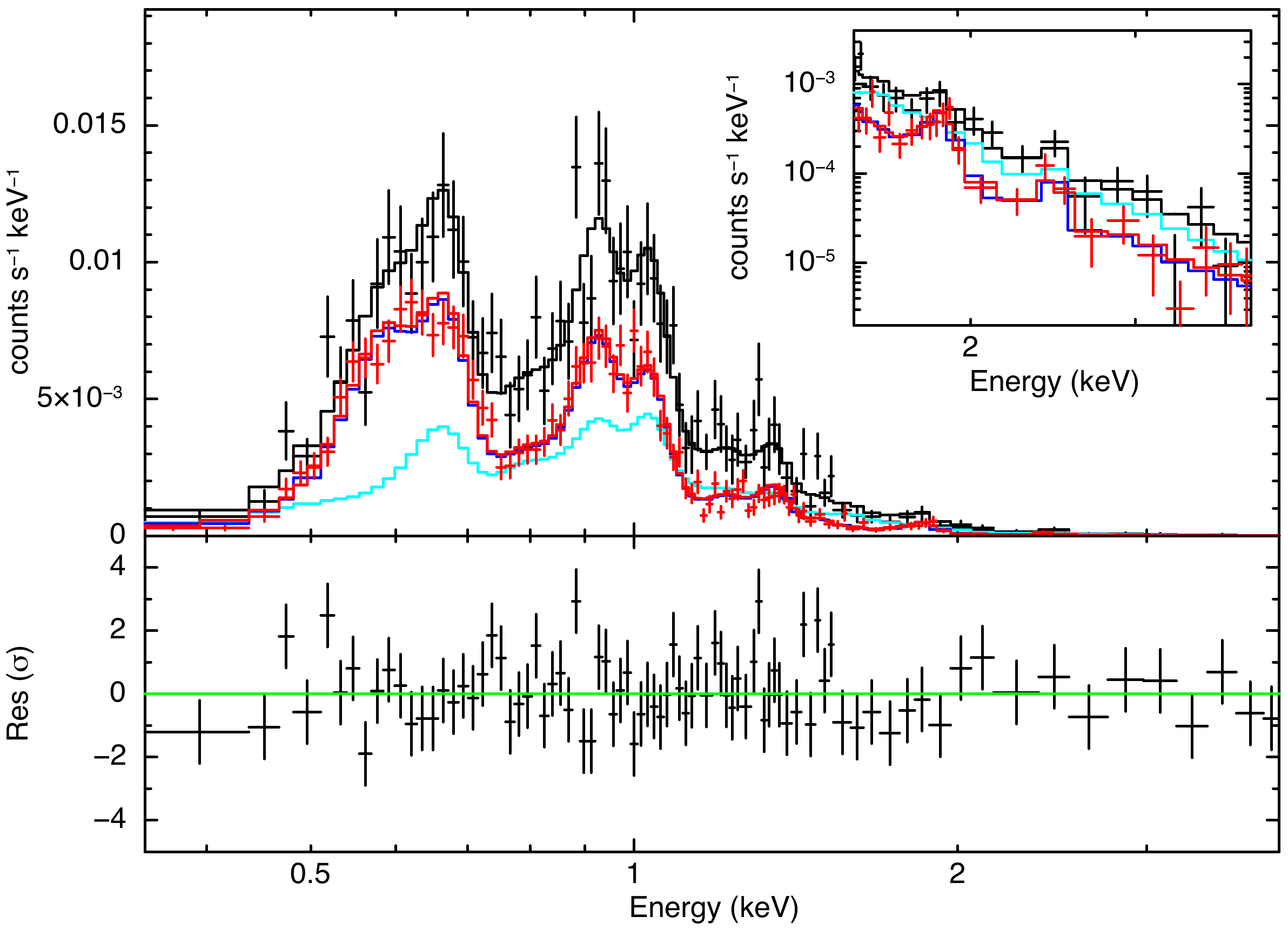}
   \caption{Source spectrum fit with a \texttt{vnei} model 
   ($n_\mathit{e}\,t \sim7.0\times10^{10}\,\mathrm{cm}^{-3}\mathrm{s}$)
   and the two sector near background model. The black points and line are the source data and model. The red points and line are the near background data and model. The cyan line is the \texttt{vnei} component in the source model. The \texttt{vnei} model parameters are listed in Table~\ref{tbl:source.vnei.nosi.nofe}.}
   \label{fig:srcvnei.nofe.2}
\end{figure}

In this section we show the spectral fits for the source spectrum with the two sector near background.
Figures~\ref{fig:scale.nearback}--\ref{fig:srcvnei.nofe.2} are the equivalent of Figures~\ref{fig:scale.alternative.nearback}--\ref{fig:srcvnei.nofe} but use the two sector background.  The plots are quite similar with the two sector background compared to the one sector background and are included for completeness.

Figure~\ref{fig:scale.nearback} shows the source spectrum fitted with the near background spectrum model, allowing only a global normalization to vary.

The source spectrum fit with the blackbody model with the temperature and normalization fixed at the values in \citet{vogt2018} is shown in Figure~\ref{fig:srcbb} for the two sector near background model, with the normalization for the near background component linked together for the source and near background spectra. 

The spectral fit with the normalization for the near background component in the source model allowed to vary independently of the normalization in the near background model, is shown in Figures~\ref{fig:srcbbfnb}. The results are listed in the third column of Table~\ref{tab:srcbb} for the two sector near background.

The spectral fit for the case in which the temperature and normalization of the blackbody component are allowed to vary and the normalization of the near background component is linked together for the source and near background spectra but also allowed to vary is shown in Figure~\ref{fig:srcbbfp}.
The results are listed in the fourth column of Table~\ref{tab:srcbb} for two sector near background.

The case in which the blackbody parameters are allowed to vary and the normalization of the near background model in the source model is allowed to vary independently of the normalization in the near background model is shown in Figure~\ref{fig:srcfbbfnb}. The results are in the fifth column of Table~\ref{tab:srcbb} for two sector near background.

Figures~\ref{fig:srcvnei.free.fe.2} and \ref{fig:srcvnei.nofe.2} show the fits for the two classes of \textit{vnei} models for the source with the two sector near background. The spectra and model fits displayed in Figure~\ref{fig:srcvnei.free.fe.2} show the model with marginally enhanced O, Ne, and S abundances with respect to SMC ISM abundances. The Si and Fe abundances are zero or close to zero. The fit results are listed in Table~\ref{tbl:source.vnei}. In Figure~\ref{fig:srcvnei.nofe.2}, the Si and Fe abundances were frozen at 0.0 and the S abundance was frozen at 0.2. The O, Ne, and Mg abundances are much lower and consistent with SMC ISM values. The ionization timescale has a significantly lower value of $n_\mathit{e}\,t\sim7.0\times10^{10}\,\mathrm{cm}^{-3}\mathrm{s}$. The fit results are shown in Table~\ref{tbl:source.vnei.nosi.nofe}.

\section{Supplemental Image Analysis Material}
\label{appdx:suppl.image.analysis}

This appendix collects the image analysis figures corresponding to combinations of point source spectrum and near background for the cases not presented in Section~\ref{sec:imganalysis}: the blackbody source spectrum with the two-sector near background, and the \texttt{vnei} source spectrum with the one- and two-sector near backgrounds. The construction of the equal-area-binning for radial profiles and the ``natural binning'' for cumulative radial profiles are described in that section.

\begin{figure*}[htb!]
  \centering
   \includegraphics[width=0.9\textwidth]
   {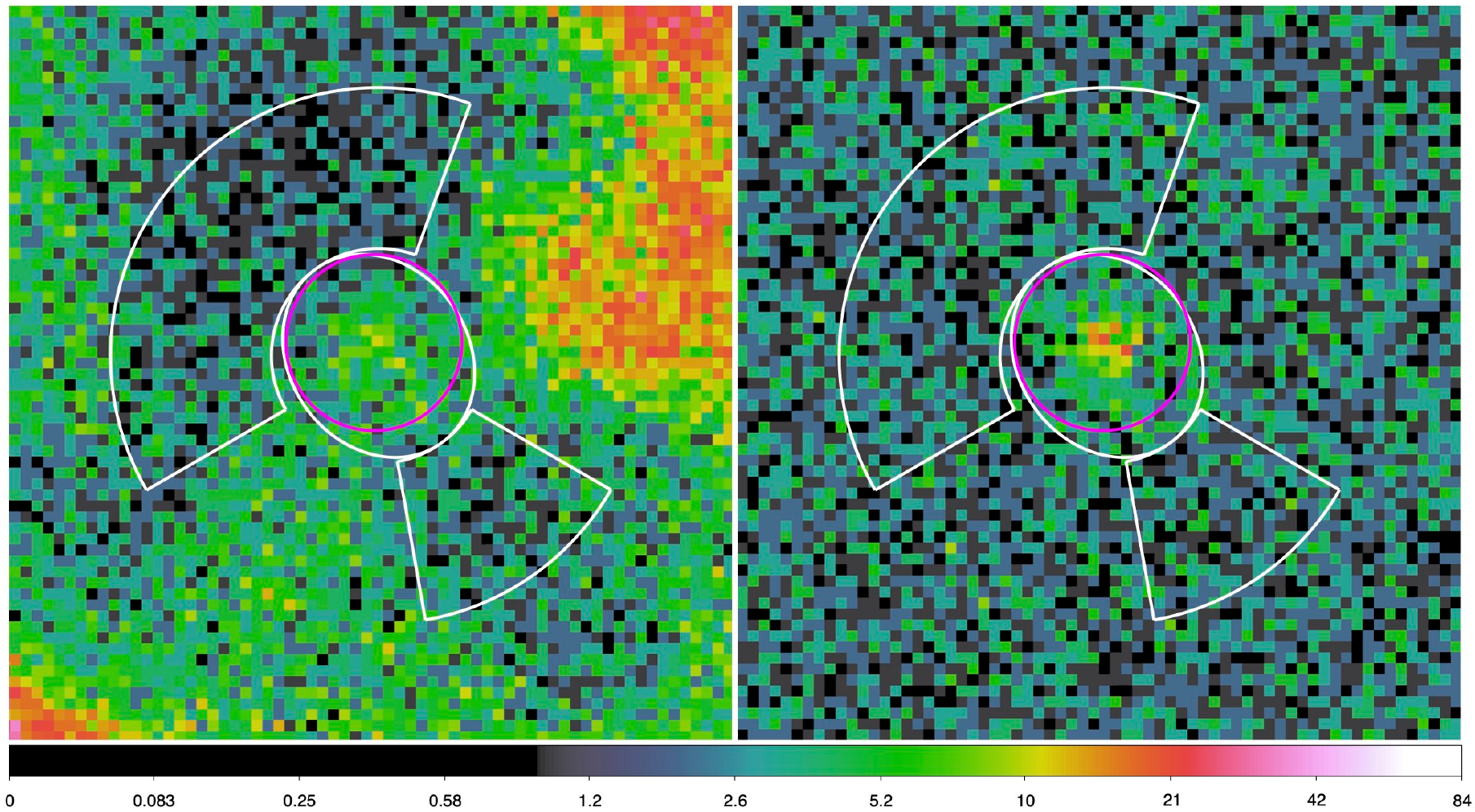}
      \caption{Left: Image of ten registered and merged observations. Right: Image created from a single instance of the 100 simulations.
      The model parameters are taken from the last column of Table~\ref{tab:srcbb}.
      The simulated blackbody point source contributes 306 counts within the source region.  The image pixel size is $0.123\arcsec$.}
   \label{fig:countsdist.twosector.bb}
\end{figure*}

\begin{figure*}[!tbh]
\begin{center}
\begin{tabular}{cc}
\includegraphics[width=0.45\textwidth]
   {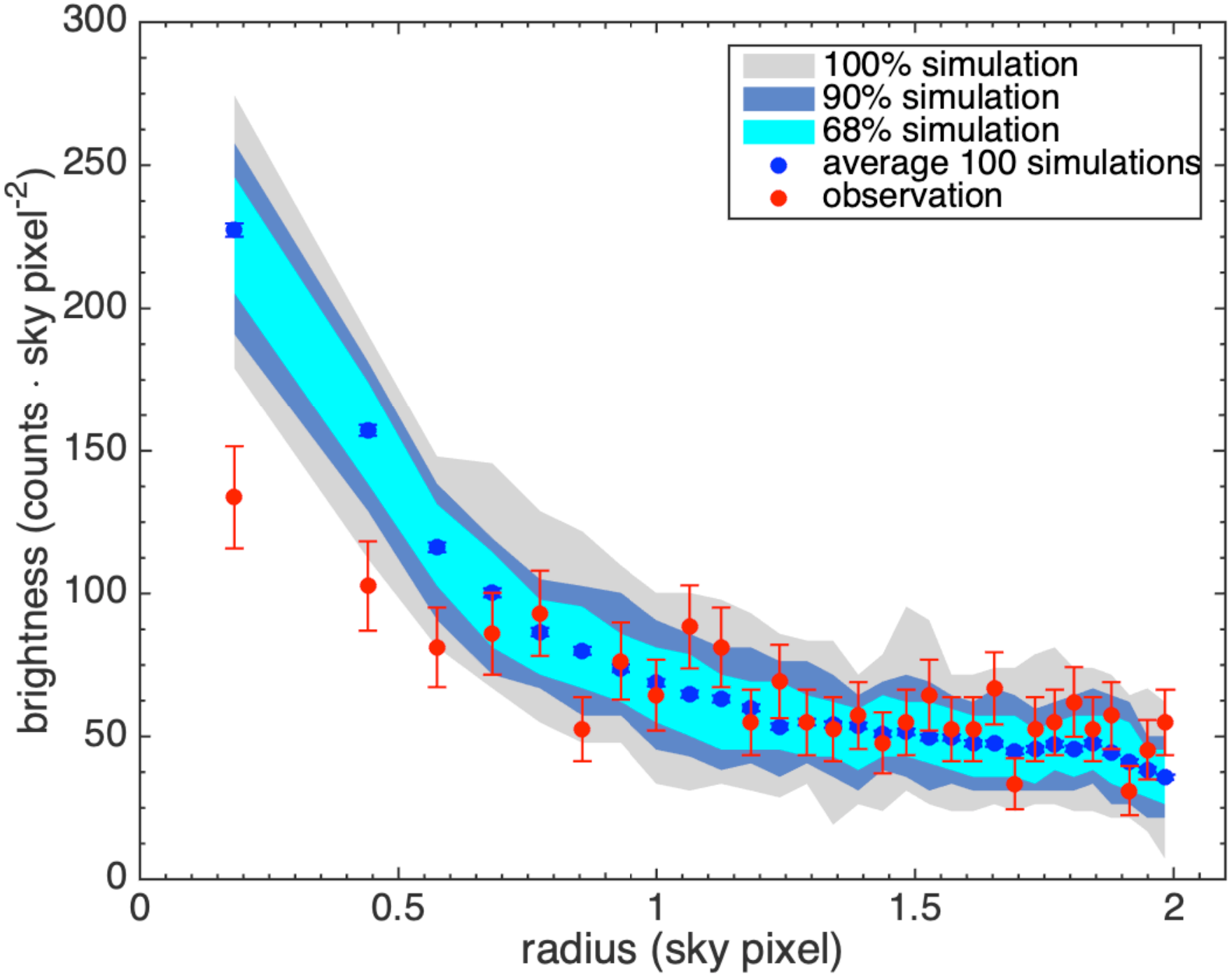} &
   \includegraphics[width=0.45\textwidth]
   {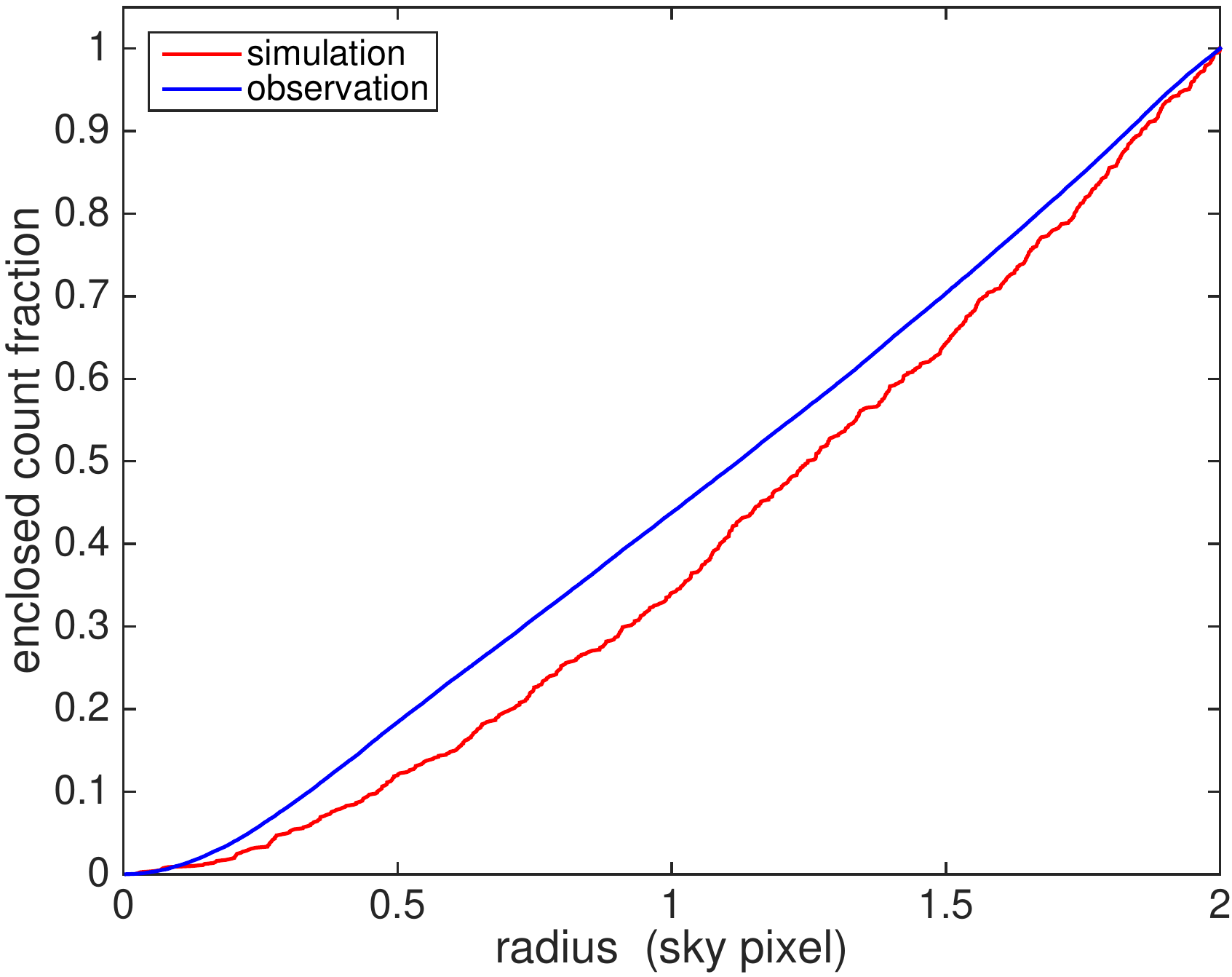}
\end{tabular}
  \caption{Surface brightness radial profiles and cumulative radial profiles (observation in magenta, simulated data in blue).  There are 306 blackbody point source counts in the source  region.
        Left: Equal-area-binned radial profile, blackbody point source and two-sector near background.
        Right: Naturally-binned cumulative surface brightness profiles normalized to one at 2.0 ACIS sky pixels (0.984\arcsec).
        Blackbody point source and two-sector near background.}
\label{fig:two.sector.profile.and.cumulative.profile.bb}
\end{center}
\end{figure*}

Figure~\ref{fig:countsdist.twosector.bb}, shows the results for the simulation  with a blackbody source ($kT$ and normalization free) and the two-sector background.
Figure~\ref{fig:two.sector.profile.and.cumulative.profile.bb} shows the equal-area-binned radial profiles for the simulation and the observation in the left panel, and the naturally binned cumulative distributions in the right panel. Table~\ref{tbl:kstest} shows the resulting Kolmogorov-Smirnov $D$ statistic and the critical $D_{0.999}(m,n)$ values. The distributions are inconsistent at the $>99.9$\% level.

\vfill\eject

\begin{figure*}[htb!]
  \centering
   \includegraphics[width=0.9\textwidth]
   {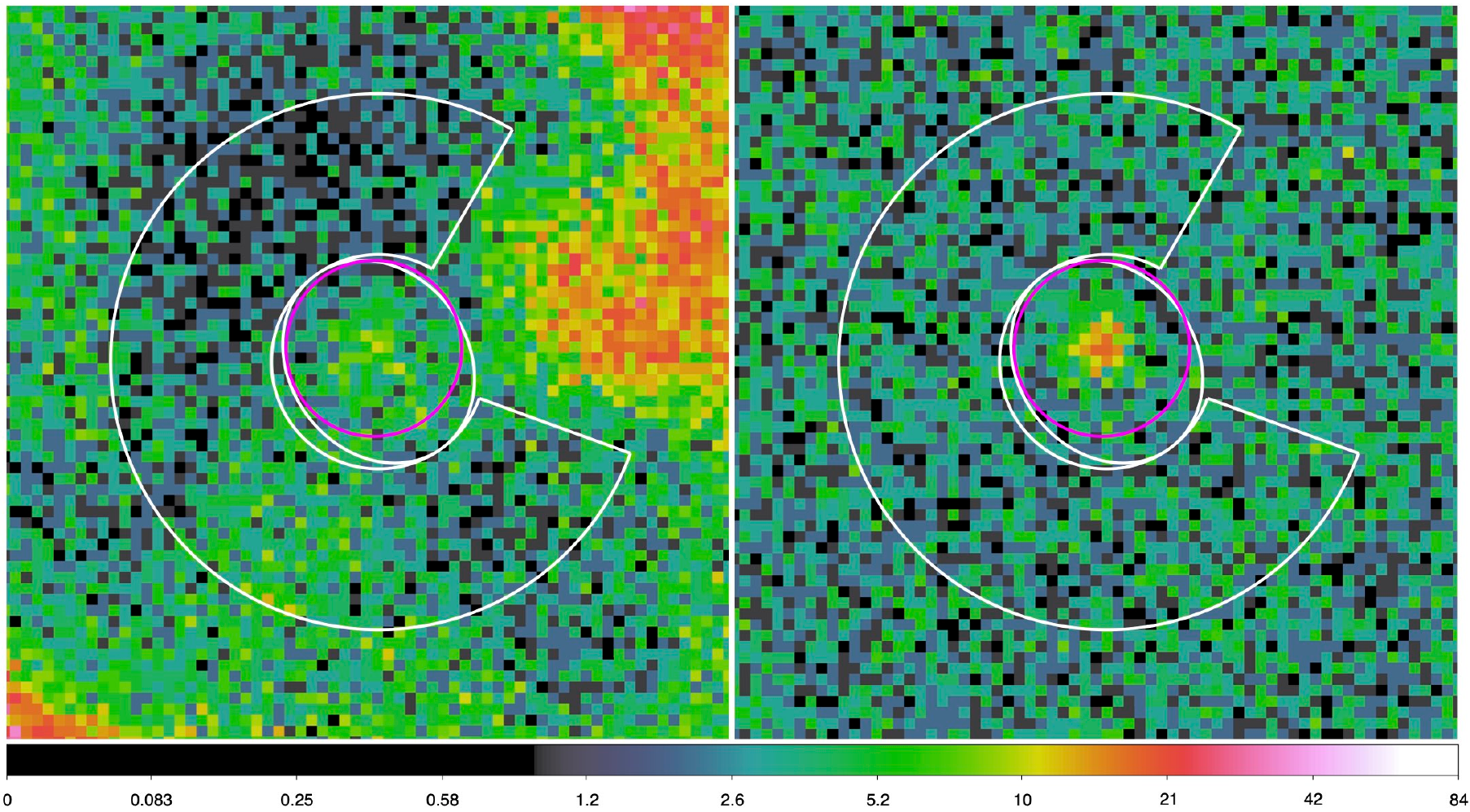}
   \caption{Left: Image of ten registered and merged observations. 
   Right: Image created from a single instance of the 100 simulations.
   The simulated \texttt{vnei} point source contributes 391  counts within the source region.  The image pixel size is $0.123\arcsec$.}
   \label{fig:countsdist.onesector.vnei}
\end{figure*}

\begin{figure*}[!tbh]
\begin{center}
\begin{tabular}{cc}
\includegraphics[width=0.45\textwidth]
   {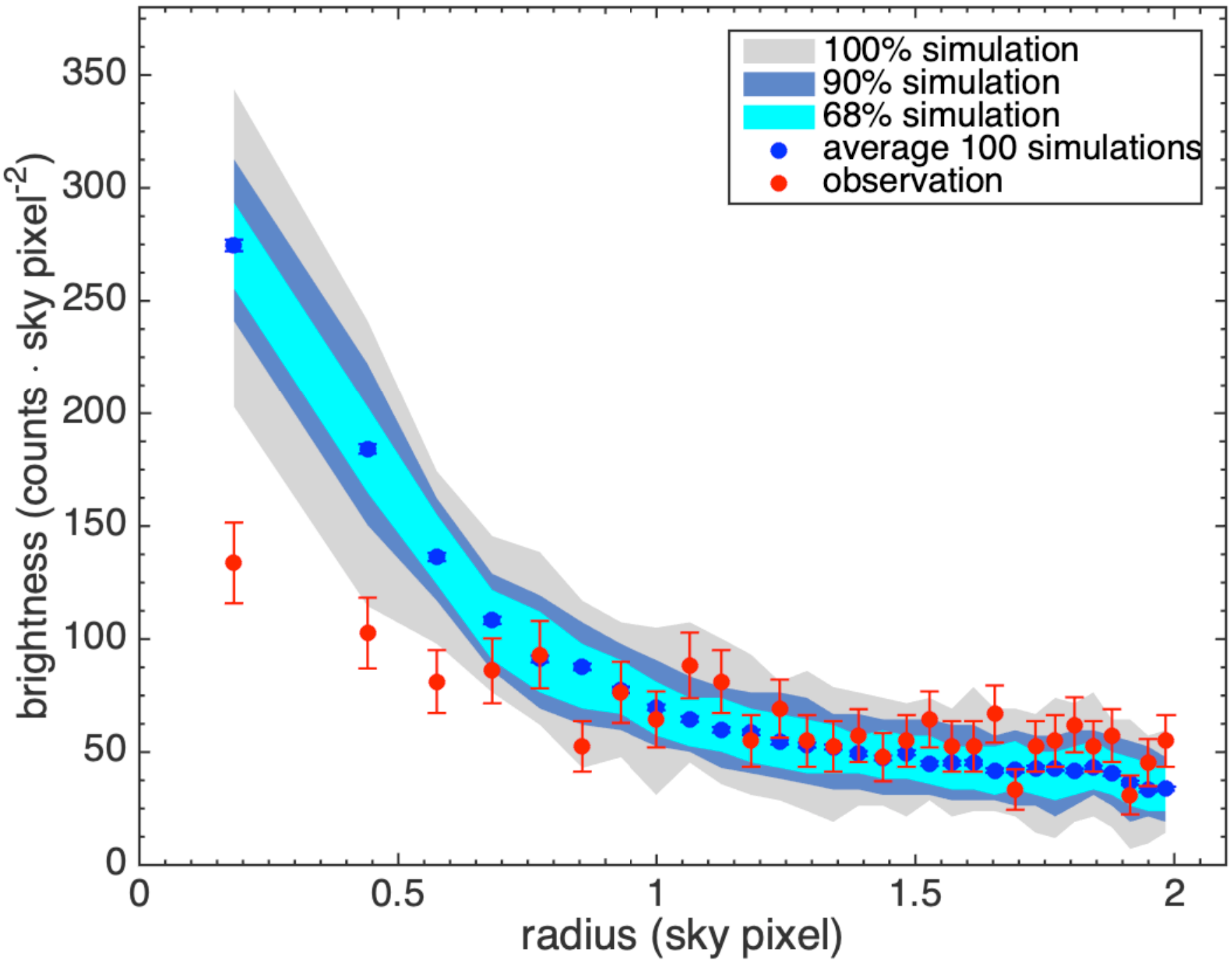} &
   \includegraphics[width=0.45\textwidth]
   {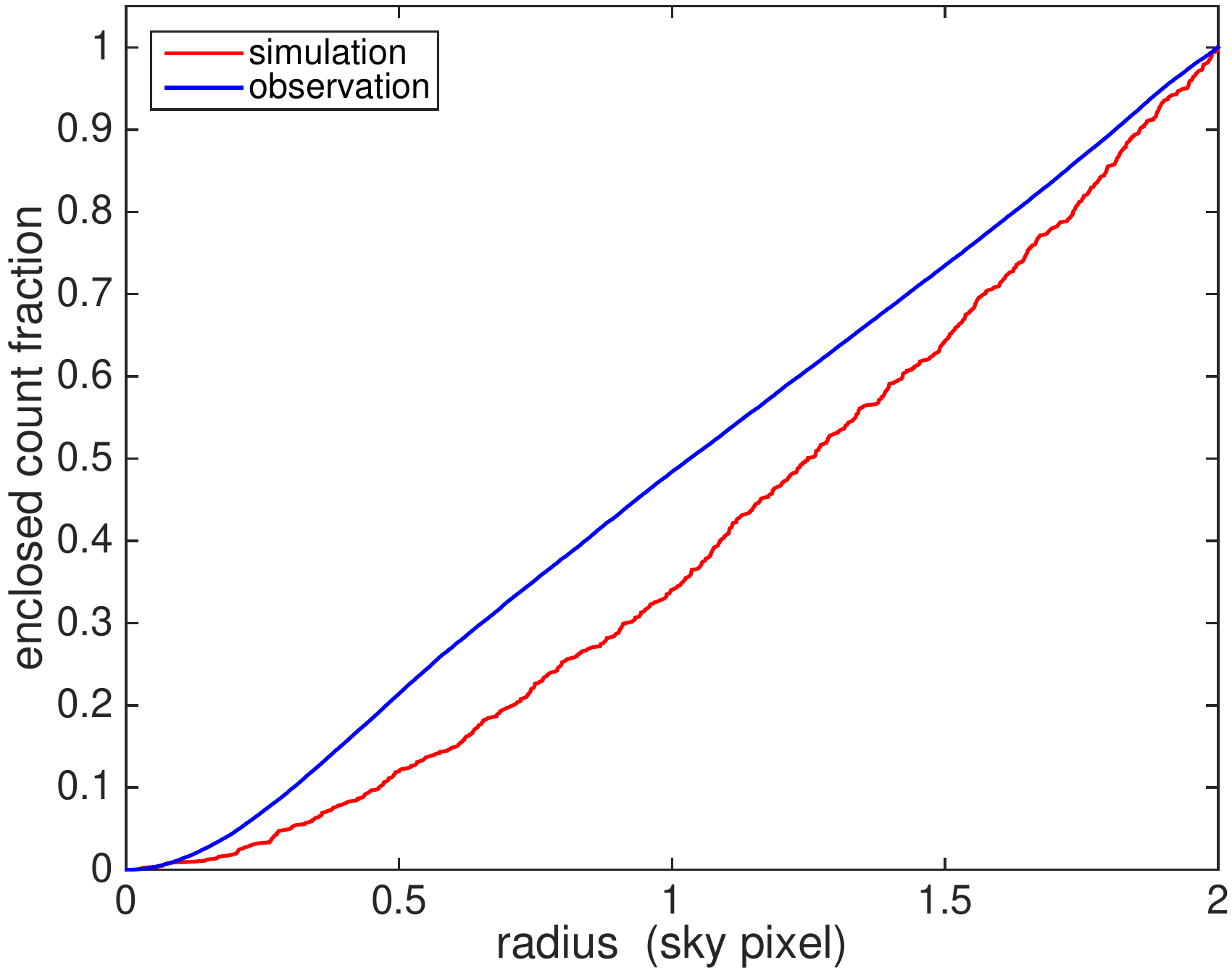}
\end{tabular}
  \caption{Surface brightness radial profiles and cumulative radial profiles (observation in magenta, simulated data in blue). There are 391 \texttt{vnei} point source counts in the source extraction region.
        Left: Equal-area-binned radial profile, \texttt{vnei} point source and one-sector near background.
        Right: Naturally-binned cumulative surface brightness profiles normalized to one at 2.0 ACIS sky pixels (0.984\arcsec).
        \texttt{vnei} point source and one-sector near background.
      }
\label{fig:one.sector.profile.and.cumulative.profile.vnei}
\end{center}
\end{figure*}

Figure~\ref{fig:countsdist.onesector.vnei}, shows the observed data (left panel) and the results for a \texttt{vnei} point source simulation with the one-sector background (right panel).
Figure~\ref{fig:one.sector.profile.and.cumulative.profile.vnei} shows the equal-area-binned radial profiles for the simulation and the observation (left panel), and the ``naturally binned'' cumulative radial distributions (right panel). Table~\ref{tbl:kstest} shows the resulting Kolmogorov-Smirnov $D$ statistic and the critical $D_{0.999}(m,n)$ values; the distributions are inconsistent at the $>99.9$\% level.

\begin{figure*}[htb!]
  \centering
   \includegraphics[width=0.9\textwidth]
   {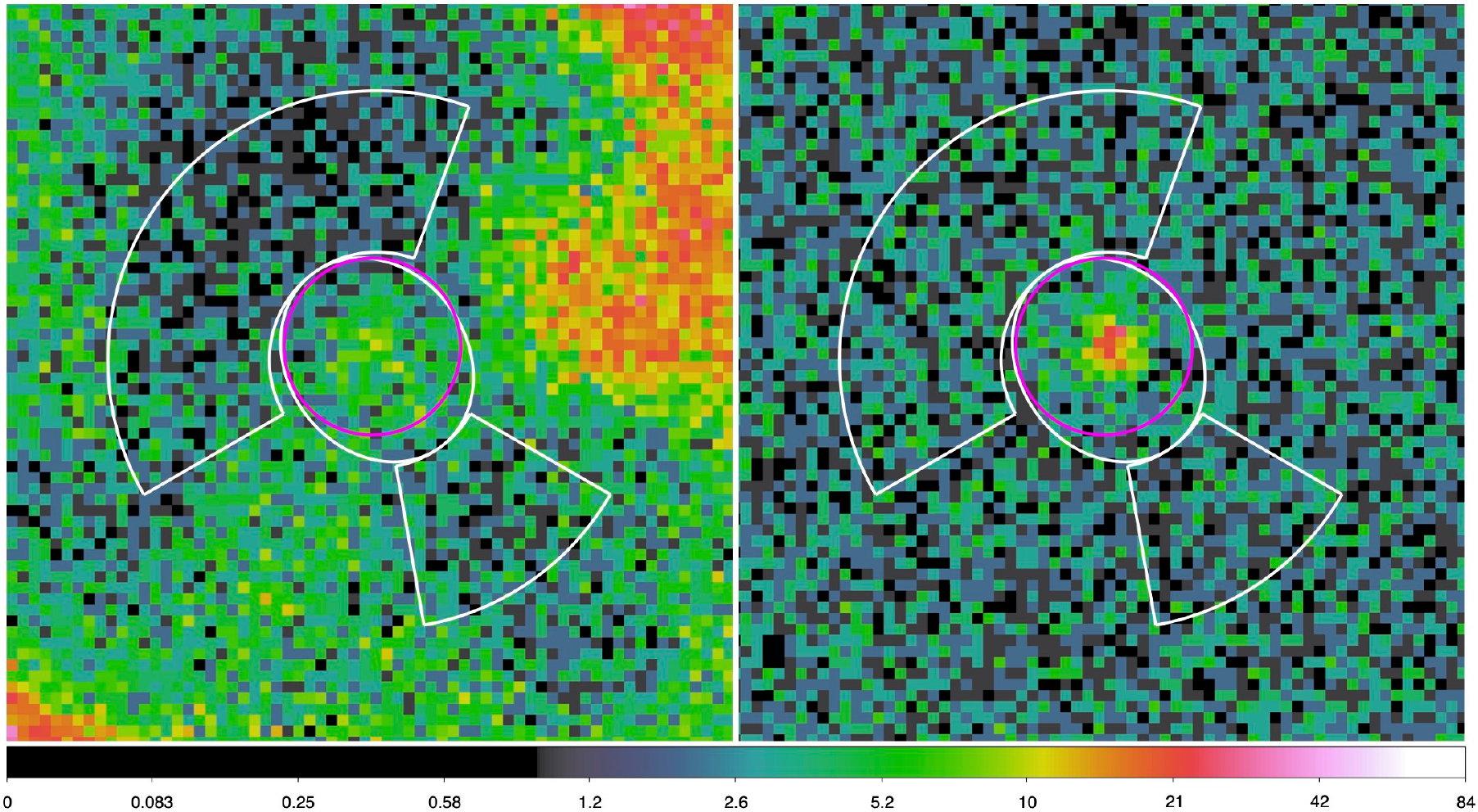}
      \caption{Left: Image of ten registered and merged observations. Right: Image created from a single instance of the 100 simulations.
      The source model is a \texttt{vnei} point source, and the corresponding two-sector near-background model was assumed.
      The simulated \texttt{vnei} point source contributes 425 counts within the source extraction region.  The image pixel size is $0.123\arcsec$.}
   \label{fig:countsdist.twosector.vnei}
\end{figure*}

\begin{figure*}[!tbh]
\begin{center}
\begin{tabular}{cc}
\includegraphics[width=0.45\textwidth]
   {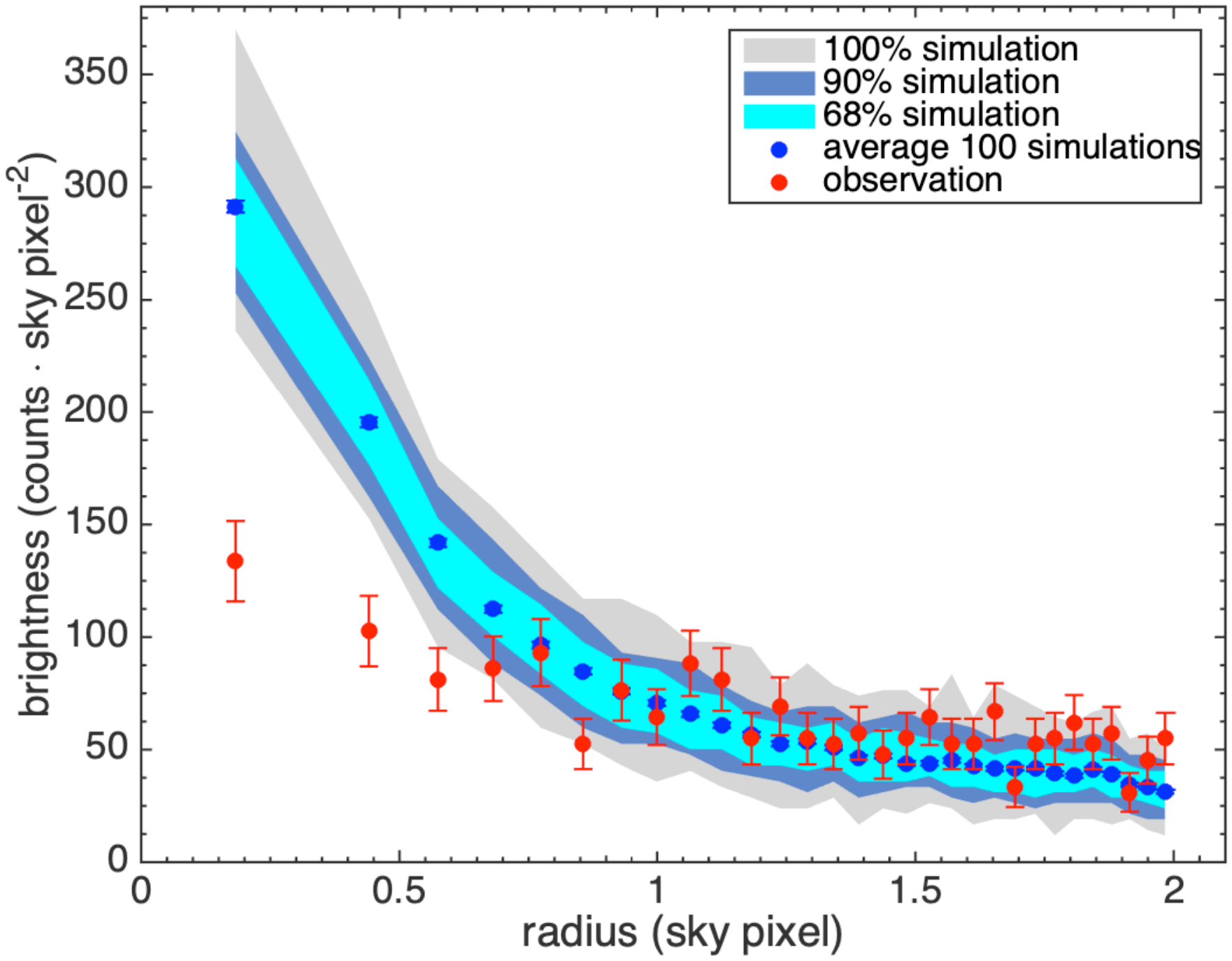} &
   \includegraphics[width=0.45\textwidth]
   {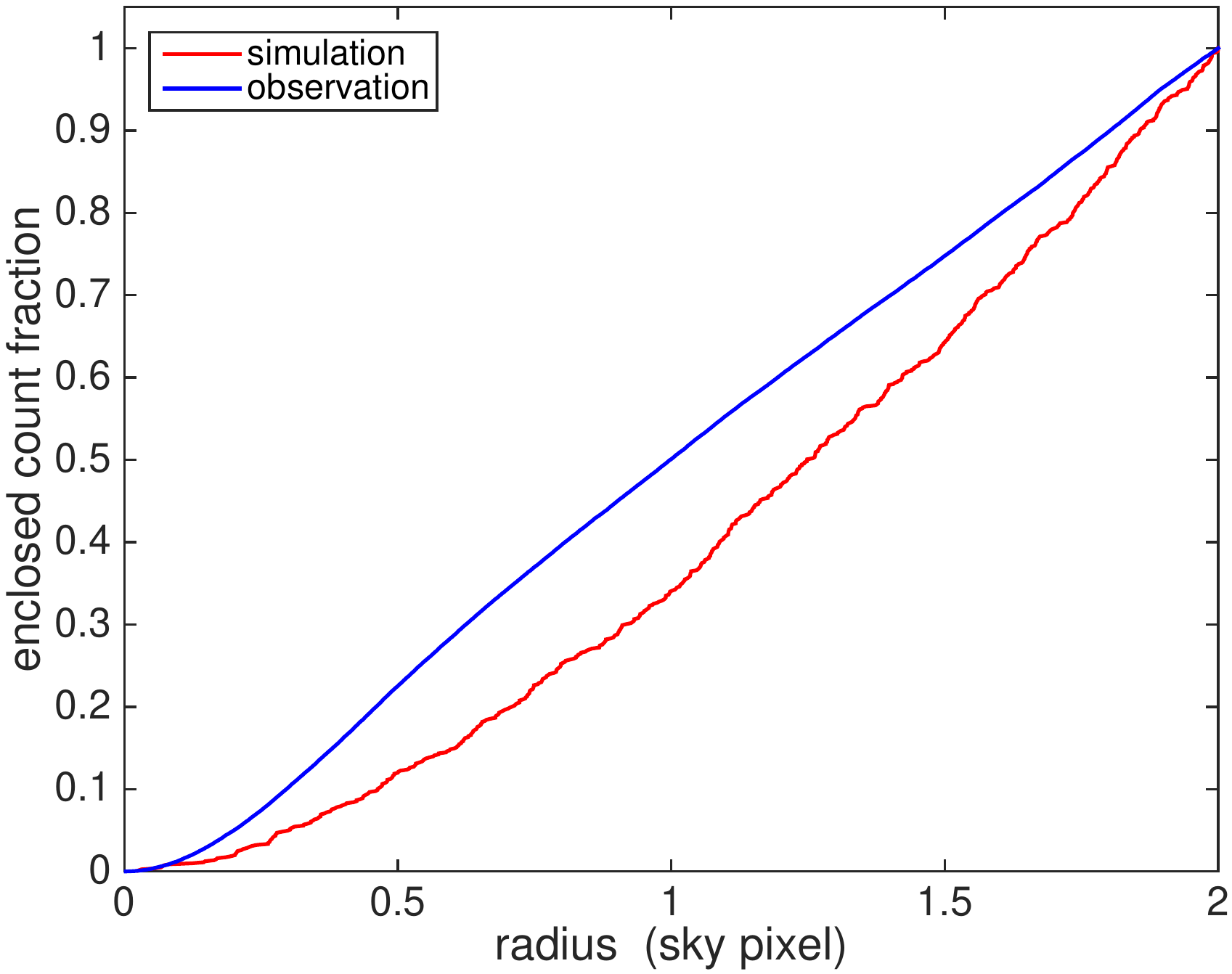}
\end{tabular}
  \caption{Surface brightness radial profiles and cumulative radial profiles (observation in magenta, simulated data in blue). The simulated \texttt{vnei} point source contributes 425 counts within the source extraction region.
        Left: Equal-area-binned radial profile, \texttt{vnei} point source and two-sector near background.
        Right: Naturally-binned cumulative surface brightness profiles normalized to one at 2.0 ACIS sky pixels (0.984\arcsec).
        \texttt{vnei} point source and two-sector near background.}
      
\label{fig:two.sector.profile.and.cumulative.profile.vnei}
\end{center}
\end{figure*}

Figure~\ref{fig:countsdist.twosector.vnei} shows the results for the simulation for a \texttt{vnei} point source with the two-sector background.
Figure~\ref{fig:two.sector.profile.and.cumulative.profile.vnei} shows the equal-area-binned radial profiles for the simulation and the observation in the left panel, and the ``naturally binned'' cumulative radial distributions in the right panel. Table~\ref{tbl:kstest} shows the resulting Kolmogorov-Smirnov $D$ statistic and the critical $D_{0.999}(m,n)$ values. The distributions are inconsistent at the $>99.9$\% level.

\vfill\eject

\vfill\eject

\end{document}